\documentclass[reprint,superscriptaddress,aps,prb]{revtex4-2}

\usepackage{times}

\usepackage{graphicx}
\usepackage{booktabs}
\usepackage{multirow}

\usepackage{hyperref}
\hypersetup{colorlinks=true,citecolor=blue,urlcolor=blue,linkcolor=blue}

\usepackage{newtxmath}
\usepackage{amsmath}
\usepackage{bm}

\usepackage[version=4]{mhchem}

\providecommand{\abs}[1]{\lvert#1\rvert}

\begin{document}

\title{Spin-Orbit Torque and Magnetization Switching in 2D Ferromagnetic Devices}

\author{Bao-Huei Huang}
\email{baohueih@gmail.com}
\affiliation{Department of Physics, National Central University, Jung-Li 32001, Taiwan}

\author{Hong Guo}
\email{hong.guo@mcgill.ca}
\affiliation{Department of Physics, McGill University, Montreal, QC, Canada H3A 2T8}

\author{Yu-Hui Tang}
\email{yhtang@cc.ncu.edu.tw}
\affiliation{Department of Physics, National Central University, Jung-Li 32001, Taiwan}

\date{\today}

\begin{abstract}
	Current-induced spin-orbit torque has emerged as a powerful technique for manipulating magnetization switching of ferromagnet/nonmagnet (FM/NM) based memory cell.
	By investigating nonequilibrium spin torque effect in a van der Waals heterobilayer, trigonal \ce{Cr3Te4}/\ce{PtTe2}, the first-principles quantum transport calculations are applied to determine both local spin induction, resulting from Rashba-Edelstein effect in the FM layer, and spin current injection, flowing from the NM to the FM layer.
	Our work reveals that local spin induction significantly generates the fieldlike torque, which primarily governs the switching current in systems with strong in-plane magnetic anisotropy.
	Our work emphasizes the importance of optimizing spin Hall effect in the NM layer for perpendicular magnetic anisotropy (PMA)-based magnetization switching and maximizing the Rashba effect in the FM layer for in-plane magnetic anisotropy (IMA)-based switching.
\end{abstract}

\maketitle

\section{Introduction}

\begin{figure}
	\centering
	\includegraphics[scale=1]{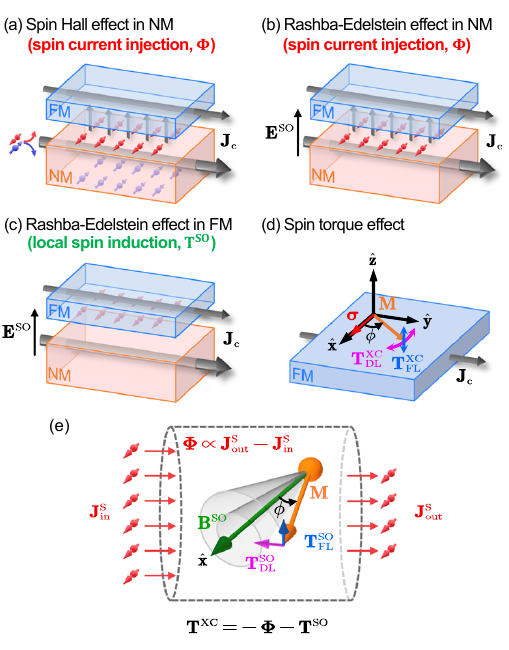}
	\caption{(a) Spin Hall effect and (b) Rashba-Edelstein effect in the NM layer induce the spin current injection from the NM to the FM layer, which is known as  \textit{spin current injection}. (c) Within the FM layer, Rashba-Edelstein effect leads to \textit{local spin induction}. (d) Both the spin current injection and local spin induction contribute to a current-induced spin density $\bm{\sigma} \parallel \mathbf{x}$ when a charge current density $\mathbf{J}_\mathrm{c} \parallel \mathbf{y}$ is applied, initiating CISOT acting on the FM magnetization in the fieldlike ($T_\mathrm{FL}$) and dampinglike ($T_\mathrm{DL}$) directions. (e) Schematic representation of the conservation relation among $\mathbf{T}^\mathrm{XC}$, $\mathbf{T}^\mathrm{SO}$ (through $\mathbf{B}^\mathrm{SO}$), and $\mathbf{\Phi}$}
	\label{fig_mechanism}
\end{figure}

The fundamental phenomenon of current-induced spin-orbit torque (CISOT) is of significant importance for efficient manipulation of magnetization in ultrathin magnetic layers and two-dimensional magnetic materials. For applications in magnetoresistive random access memory (MRAM), devices based on CISOT overcome the shortcomings of high writing current density and reading disturbance associated with the current-driven spin-transfer torque (STT) \cite{Ramaswamy2018,Manchon2019}. In typical ferromagnet/nonmagnet (FM/NM) heterobilayers, there are two primary spin-orbit coupling (SOC) mechanisms that contribute to the generation of nonequilibrium spin density when an in-plane charge current flows through: the bulk spin Hall effect and the interfacial Rashba-Edelstein effect.
For the bulk spin Hall effect [see Fig.~\ref{fig_mechanism}(a)], the NM layer, typically using heavy metals with strong bulk SOC, deflects oppositely oriented spins in opposite directions, thereby generating a transverse spin current. The orientation of this spin current is determined by the sign of the spin Hall conductivity and the direction of the applied charge current. Examples include Co/Pt and CoFeB/Ta heterostructures \cite{Sinova2004,Sinova2015,Liu2012,Liu2012Ralph,Garello2013}.
On the other hand, interfacial SOC (ISOC) can induce Rashba spin-splitting at the interface between the FM and NM layers. When an in-plane charge current flows through the NM layer [see Fig.~\ref{fig_mechanism}(b)] or the FM layer [see Fig.~\ref{fig_mechanism}(c)], a nonequilibrium spin density is generated via the Rashba-Edelstein charge-to-spin conversion process \cite{Edelstein1990,Miron2011,Lee2015,Nakayama2016,Amin2020}.

Relatively recently, many new two-dimensional ferromagnetic (2DFM) van der Waals (vdW) materials have been discovered which provide a broad material phase space for the development of next-generation spintronic devices. These 2DFM materials include \ce{CrI3} and \ce{Fe3GeTe2}, with Curie temperatures ($T_c$) of 45 and 225 K, respectively \cite{Huang2017,Deng2018,Fei2018,Alghamdi2019}. Notably, \ce{CrI3} exhibits a transition from an antiferromagnetic to a ferromagnetic state depending on the layer count. Higher $T_c$ 2DFM materials have been experimentally reported in Cr-Te families, ranging from 165 to 295 K for trigonal \ce{Cr2Te3} \cite{Li2019} and approximately 322 or 344 K for monoclinic \ce{Cr3Te4} (m-\ce{Cr3Te4}) \cite{Yamaguchi1972,Chua2021,Goswami2024,Hideki2024}.
%
Spintronic devices using all-vdW materials, \ce{CrI3}/\ce{TaSe2} and \ce{Fe3GeTe2}/\ce{WTe2}, have been extensively proposed in theoretical and experimental literatures \cite{Dolui2020,Shin2022}. In these heterostructures, current-induced spin densities, generated by either the spin Hall effect [Fig.~\ref{fig_mechanism}(a)] or the Rashba-Edelstein effect [Fig.~\ref{fig_mechanism}(b)], flow from the NM layer into the FM layer, a process known as spin current injection, initiating spin torques on the FM magnetization. Conversely, the spin density generated within the FM layer [Fig.~\ref{fig_mechanism}(c)], referred to as local spin induction, also contributes to the spin torques.
Both spin current injection and local spin induction provide fundamental mechanisms for initiating CISOT on the FM magnetization, either individually or simultaneously [Fig.~\ref{fig_mechanism}(d)] \cite{Haney2013,Mahfouzi2020}. However, distinguishing the contributions to CISOT by each mechanism remains a significant challenge.

As CISOT can be decomposed into the fieldlike torque (FLT, $T_\mathrm{FL}$) and the dampinglike torque (DLT, $T_\mathrm{DL}$) components [Fig.~\ref{fig_mechanism}(d)], several experiments have demonstrated that $T_\mathrm{DL}$ is generally larger than $T_\mathrm{FL}$. In the experiment involving \ce{Py/PtTe2}, the $T_\mathrm{DL}$ efficiency was observed to range from $0.058$ to $0.152$, while the $T_\mathrm{FL}$ efficiency ranged from $-0.002$ to $-0.004$ \cite{Xu2020}. For \ce{Py/WS2}, $T_\mathrm{FL}/T_\mathrm{DL} \approx 0.17$ is reported \cite{Lv2018}.
A recent theoretical toy model by Kim \textit{et al}. \cite{Kim2017} indicates that a charge current inside the NM layer generates a nonequilibrium spin current flowing into the FM layer, exerting both $T_\mathrm{FL}$ and $T_\mathrm{DL}$ on the magnetization with approximately the same order of magnitudes. In contrast, when the current flows in the FM layer, the finite magnitude of exchange splitting in the FM layer and the resulting evanescent states produce a larger $T_\mathrm{FL}$.
While CISOT has received extensive theoretical and experimentally investigations, the origin and interplay between $T_\mathrm{FL}$ and $T_\mathrm{DL}$, as well as their respective roles in spin current injection and local spin induction, remain unclear. Furthermore, the impact of these components in CISOT on the switching current and their consequent effects on magnetization dynamics require further investigation.

\begin{figure}
	\centering
	\includegraphics[scale=1]{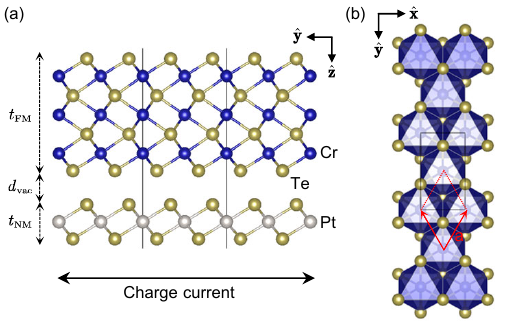}
	\caption{(a) Side view and (b) top view of the central region \ce{Cr3Te4}/\ce{PtTe2} heterobilayer, where both the \ce{Cr3Te4} and \ce{PtTe2} monolayers possess $P\bar{3}m1$ space group. The optimized structure parameters are provided in Table~\ref{tab_struct_params}. The red box denotes the primitive cell with lattice constant $a$. For nonequilibrium calculations, the charge current is applied along the $y$ direction.}
	\label{fig_struct_textures}
\end{figure}

In this article, we theoretically investigate the roles of spin current injection and local spin induction in the fieldlike $T_\mathrm{FL}$ and the dampinglike $T_\mathrm{DL}$ components in the CISOT effect, using the all-vdW \ce{Cr3Te4}/\ce{PtTe2} magnetic heterobilayer as a prototypical device structure, shown in Fig.~\ref{fig_struct_textures}. The trigonal \ce{Cr3Te4} (t-\ce{Cr3Te4}) monolayer which was predicted to be a vdW-2DFM with a high $T_c$ of up to 411 K by data science and density functional theory (DFT) calculations \cite{Zhu2018}, is chosen as the FM layer.
Using first-principles quantum transport calculations, we find that local spin induction significantly contributes to $T_\mathrm{FL}$, while spin current injection contributes to $T_\mathrm{DL}$.
Macrospin simulations reveal that in \ce{Cr3Te4}/\ce{PtTe2}, which exhibits in-plane magnetic anisotropy (IMA), $T_\mathrm{FL}$ arising from the local spin induction dominates the switching current.
Our results suggest that for efficient IMA switching, enhancing the Rashba effect in the FM layer is key to boosting $T_\mathrm{FL}$. In contrast, for perpendicular magnetic anisotropy (PMA) switching, optimizing spin current injection, typically via the spin Hall effect, is essential.

\section{Distinguishing spin current injection and local spin induction}

\begin{table*}[!t]
	\centering
	\caption{The optimized structural parameters, magnetic moments, and magnetizations for the \ce{Cr3Te4} (FM) monolayer, \ce{PtTe2} (NM) monolayer, and \ce{Cr3Te4}/\ce{PtTe2} (hetero) heterobilayer. $a$ and $t$ represent the lattice constant and the thickness, respectively. $d_\text{vac}$ is the interlayer distance between the \ce{Cr3Te4} and \ce{PtTe2} monolayers. $m_\text{FM}$ and $M_\text{FM}$ are the magnetic moments and magnetizations for the \ce{Cr3Te4} monolayer, respectively. $N_\text{Cr}$ is the number of Cr atoms in the unit cell.}
	\begin{ruledtabular}
		\begin{tabular*}{\textwidth}{@{\extracolsep{\fill}}ccccccc}
			Structure & $a$ ({\AA}) & $t_\text{NM}$ ({\AA}) & $d_\text{vac}$ ({\AA}) & $ t_\text{FM}$ ({\AA}) & $m_\text{FM}$ ($\mu_\text{B}/N_\text{Cr}$) & $M_\text{FM}$ (A/m) \\
			\midrule
			\ce{Cr3Te4}                    & $4.05$ & -           & -       & $9.60$ & $3.42$ & $684 \times 10^3$ \\
			\ce{PtTe2}                       & $3.98$ & $2.82$ & -      & -      & $0.0$  & $0.0 $            \\
			\ce{Cr3Te4}/\ce{PtTe2} & $4.05$ & $2.74$ & $2.60$ & $9.66$ & $3.50$ & $708 \times 10^3$ \\
		\end{tabular*}
	\end{ruledtabular}
	\label{tab_struct_params}
\end{table*}

%
To identify the underlying physical mechanisms of CISOT, we employ the continuity equation for the spin density in the FM layer, ${\bf S}_\mathrm{FM}$.
Applying the Heisenberg equation of motion, this is expressed as \cite{Huang2023b}
\begin{equation}
	\frac{d {\bf S}_\mathrm{FM}}{d t} = \frac{1}{i\hbar} \left< \left[ \hat{\bf S}_\mathrm{FM}, \hat{\mathcal{H}} \right] \right> = {\bf \Phi}_\mathrm{FM} + {\bf T}_\mathrm{FM}^{\mathrm{XC}} + {\bf T}_\mathrm{FM}^{\mathrm{SO}},
\end{equation}
where $\hat{\bf S}_\mathrm{FM}$ is the vector of spin operators projected to the FM layer, and $\hat{\mathcal{H}}$ represents the system Hamiltonian. The terms ${\bf \Phi}_\mathrm{FM}$, ${\bf T}_\mathrm{FM}^{\mathrm{XC}}$ and ${\bf T}_\mathrm{FM}^{\mathrm{SO}}$ correspond to the spin current accumulation, exchange spin torque and spin-orbit torque, respectively.
As depicted in Fig.~\ref{fig_mechanism}(d), ${\bf T}_\mathrm{FM}^{\mathrm{XC}}$ is the spin torque directly acting on the magnetization ${\bf M}$.
The spin current accumulation, $\mathbf{\Phi}_\mathrm{FM} \propto \mathbf{J}^\mathrm{S}_\mathrm{out} - \mathbf{J}^\mathrm{S}_\mathrm{in}$ sketched in Fig.~\ref{fig_mechanism}(e), accounts for the net flow of spin current into ($\mathbf{J}^\mathrm{S}_\mathrm{in}$) and out of ($\mathbf{J}^\mathrm{S}_\mathrm{out}$) the FM layer region.
$\mathbf{\Phi}_\mathrm{FM}$ thus in turn encompasses the spin current injection flowing from the NM layer to the FM layer, either from the spin Hall effect or the Rashba-Edelstein effect [Figs.~\ref{fig_mechanism}(a) and \ref{fig_mechanism}(b)].
%
Additionally, the current-induced spin density within the FM layer, attributed to the Rashba-Edelstein effect [Fig.~\ref{fig_mechanism}(c)], is examined through ${\mathbf{T}}^\mathrm{SO}_\mathrm{FM}$, which arises from an effective SOC field ${\bf B}^\mathrm{SO}$ within the FM layer as shown in Fig.~\ref{fig_mechanism}(e).
%
In a steady state (${d {\bf S}_\mathrm{FM}}/{d t}=0$), ${\bf T}_\mathrm{FM}^{\mathrm{XC}}$ can be related to the spin current accumulation and the spin-orbit torque \cite{Go2020,Dolui2020,Huang2023b}:
\begin{equation}\label{Txc}
	{\bf T}_\mathrm{FM}^{\mathrm{XC}} = -{\bf \Phi}_\mathrm{FM} - {\bf T}_\mathrm{FM}^{\mathrm{SO}} .
\end{equation}
Namely, investigating the constituents of ${\bf \Phi}_\mathrm{FM}$ and ${\bf T}_\mathrm{FM}^{\mathrm{SO}}$ allows us to effectively distinguish between the spin current injection and local spin induction, respectively.

We note that, in a strict real-space representation, the Rashba-Edelstein effect cannot be uniquely determined to either the NM or FM layer [Fig.~\ref{fig_mechanism}(b) and \ref{fig_mechanism}(c)], since electron density and wavefunctions naturally extend across the FM/NM interface.
However, our DFT calculations employ localized atomic orbital (LAO) basis sets directly, in which the Hamiltonian and related quantities are orbital-resolved. Since LAOs are grouped by atomic sites, we can decompose the calculated spin torques into contributions from atoms belonging to the NM or FM region.
This orbital-resolved decomposition is explicitly preserved in our derivation of spin torques \cite{Huang2023b}, which differs from the approaches involving real-space wavefunctions \cite{Freimuth2014,Dolui2020}.
While this decomposition does not alter the fact that the Rashba-Edelstein effect originates from interfacial inversion symmetry breaking, it provides a practical way to identify which atomic sites contribute most strongly to the effect.

\section{Computational details}

For structure optimization, first-principles calculations using the \textsc{VASP} package based on density functional theory (DFT) \cite{Kresse1996,Kresse1999} are employed using projector augmented wave (PAW) pseudopotential and GGA based PBE exchange-correlation (XC) functional \cite{Perdew1996}. The plane-wave cutoff energy is 500 eV, and the $k$-space mesh is $15 \times 15 \times 1$. To account for electron correlation effects in the Cr-$3d$ orbitals, the GGA$+U$ method is employed. A Hubbard $U$ value of 4.0 eV for Cr is determined using the linear response approach \cite{Cococcioni2005}.
The convergence criteria for the electronic self-consistent loop and for the ionic relaxation loop are the total/band energy less than ${10}^{-6}$ eV and the atomic forces less than ${10}^{-2}$ eV/{\AA}, respectively. To accurately optimize the interlayer distance between \ce{Cr3Te4} and \ce{PtTe2}, the nonlocal van der Waals density functional based on the optB86b-vdW \cite{Klime2011} method is included.
Since the difference of the in-plane lattice constants between \ce{Cr3Te4} and \ce{PtTe2} is small (with lattice constants $a_\text{FM}=4.05$ {\AA} for \ce{Cr3Te4} and $a_\text{NM}=3.98$ {\AA} for \ce{PtTe2}), the bilayers are optimized using a primitive cell scheme. The out-of-plane cell length is fixed at $c=40$ Å to include a substantial vacuum layer, preventing interactions between periodic cells. The optimized in-plane lattice constant is determined to be $a_\text{hetero}=4.05$ Å. The optimized structural parameters and magnetizations are listed in Table~\ref{tab_struct_params}.
Notably, the maximum lattice constant strain is $(a_\text{NM}-a_\text{hetero})/a_\text{hetero} \approx -1.73\%$.

Using the optimized structure, we compute the self-consistent Hamiltonian of two-probe devices from first-principles with the \textsc{NanoDCAL} package \cite{Taylor2001,Waldron2007,Ke2008}. This approach is based on nonequilibrium Green's function (NEGF) combined with DFT.
The two-probe devices are constructed with the length of electrode and central scattering region being 7.02 and 21.05 {\AA}, respectively.
We employ a double-$\zeta$ plus double-polarization (DZP) localized atomic orbital (LAO) basis set and an LDA-based PZ XC functional \cite{Perdew1981} (see also Appendix~\ref{sec_note_lda}).
The parameters for these calculations include a cutoff energy of 50 Hartree for real space mesh grids and a cutoff length of 40 Bohr for $k$-space sampling in the scattering region. For semi-infinite electrodes, these values are increased to 150 Hartree and 80 Bohr, respectively. The convergence criteria for both the self-consistent Hamiltonian and the total energy are ${10}^{-5}$ eV.
For each rotated angle of the spin moments in \ce{Cr3Te4}, the Hamiltonians are computed self-consistently in the presence of spin-orbit coupling (SOC) and an applied nonequilibrium electric bias, $V_b$. The voltages applied to the left and right electrodes along the $y$ direction are $V_L=V_b/2$ and $V_R=-V_b/2$, respectively. Note that the Hubbard $U$ correction is not included in these NEGF-DFT calculations due to convergence challenges under nonequilibrium conditions and because its effect on SOC-driven transport remains complex and system-dependent.
Also, the current results are obtained in the clean limit, where explicit structural or chemical disorder is omitted.
%

Finally, we calculate the angular and current dependence of CISOT acting on the magnetization of FM \ce{Cr3Te4} using our \textsc{JunPy} package employing the NEGF method \cite{Huang2023a,Huang2023b}.
We use the $k$-space sampling set at $50\times1\times1$ without time-reversal symmetry and an imaginary energy broadening of 0.002 eV in the retarded Green's function.
Importantly, because our system is modeled as an open device with semi-infinite electrodes extending along the $y$ axis, transport observables are evaluated exclusively within the isolated central scattering region to properly account for the multidimensional spin flux.

\section{Results and Discussion}

\subsection{ISOC-induced spin textures}
\label{sec_spin_textures}

\begin{figure}
	\centering
	\includegraphics[scale=1]{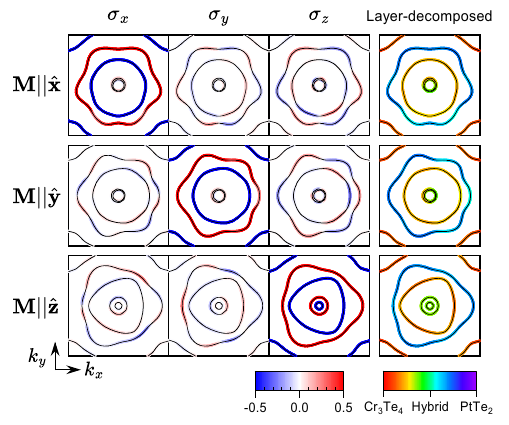}
	\caption{ISOC-induced spin textures by projecting onto the Pauli matrices ($\sigma_x$, $\sigma_y$, and $\sigma_z$) near the $\Gamma$-point at Fermi surface with $E_\mathrm{F}=0.0$ eV. The magnetization direction of \ce{Cr3Te4} is aligned along various directions. For the layer-decomposed analysis, the hybridization of \ce{Cr3Te4} and \ce{PtTe2} (light-blue color) results in the the bands that causes Rashba spin-splitting.}
	\label{fig_spin_textures_2}
\end{figure}

%
We first demonstrate the expectation value of spin density, $\sigma_{x}$, $\sigma_{y}$, and $\sigma_{z}$, of \ce{Cr3Te4}/\ce{PtTe2} at equilibrium ($V_b=0$) to characterize the ISOC effect.
Figure \ref{fig_spin_textures_2} displays the spin textures at the Fermi surface ($E_\mathrm{F} = 0$ eV) near the $\Gamma$-point with SOC. The magnetization direction of \ce{Cr3Te4} monolayer is aligned along the $x$, $y$, or $z$ axis ($\mathbf{M} \parallel \hat{\bf x}$, $\mathbf{M} \parallel \hat{\bf y}$, or $\mathbf{M} \parallel \hat{\bf z}$). Additional details regarding the band structures and the entire Brillouin zone are provided in Appendix~\ref{sec_detail_bands}.
As shown in the layer-decomposed plot of Fig.~\ref{fig_spin_textures_2}, the hybridization between \ce{Cr3Te4} and \ce{PtTe2} (light-orange and light-blue colors) leads to the formation of two hybrid-bands near the $\Gamma$-point, on which the Rashba spin-splitting is observed.
For $\mathbf{M} \parallel \hat{\bf y}$, $\sigma_x$ exhibits inversion symmetry along the $k_y$ direction, namely $\sigma_x (k_y)=-\sigma_x (-k_y)$, with nonzero $\sigma_x$ values across $k_y$, indicating a spin direction perpendicular to the electron momentum. A similar behavior is observed for $\sigma_y$ along the $k_x$ direction for $\mathbf{M} \parallel \hat{\bf x}$.
These characteristics of spin-splitting confirm the presence of ISOC induced by the proximity of \ce{PtTe2} capping layer.
%

\subsection{Current-induced spin-orbit torque}
\label{sec_cisot}

\begin{figure}
	\centering
	\includegraphics[scale=1.2]{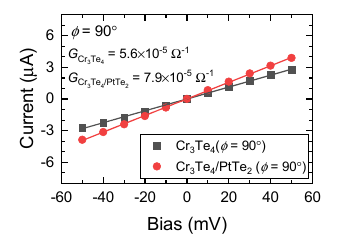}
	\caption{Current-voltage characteristics for both the \ce{Cr3Te4} monolayer and the \ce{Cr3Te4}/\ce{PtTe2} heterobilayer, where $G$ is the conductance.}
	\label{fig_ivcurve}
\end{figure}
We drive a charge current density $\mathbf{J}_\mathrm{c}$ along the $y$ direction by applying a voltage bias $V_b$, as sketched in Fig.~\ref{fig_mechanism}(d).
%
Figure~\ref{fig_ivcurve} shows the current-voltage characteristics, where the current is computed using the Landauer-Büttiker formula \cite{Datta1995}. The similar conductance between the \ce{Cr3Te4} monolayer ($5.6 \times 10^{-5}\ \mathrm{\Omega^{-1}}$) and the \ce{Cr3Te4}/\ce{PtTe2} heterobilayer ($7.9 \times 10^{-5}\ \mathrm{\Omega^{-1}}$) suggests an assumption that the charge current fully flows through the \ce{Cr3Te4} layer.
Since the presence of ISOC provides an effective electric field $E^\mathrm{SO}$ along the $z$ direction, the flowing electrons within the FM layer is subjected to an effective magnetic field, ${\bf B}^\mathrm{SO} \propto E^\mathrm{SO}\hat{\mathbf{z}} \times  \mathbf{J}_\mathrm{c}$, along the $x$ direction.

Note that we do not consider the cases with current applied along the $x$ direction. As shown by the $\sigma_z$ spin textures in Fig.~\ref{fig_spin_textures_2}, the band structure exhibits inversion symmetry along $k_x$ for arbitrary orientations of $\mathbf{M}$. This symmetry implies that an $x$-directed current can generate an additional $z$-polarized spin accumulation. In such cases, the system would generate two distinct current-induced spin-orbit fields, one along $\hat{\bf y}$ and another along $\hat{\bf z}$, both contributing to the CISOT depending on the current direction. In this work, we restrict our analysis to the simpler geometry where the current flows along $y$, corresponding to the conventional Rashba-Edelstein scenario.

\begin{figure}[!t]
	\centering
	\includegraphics[scale=1]{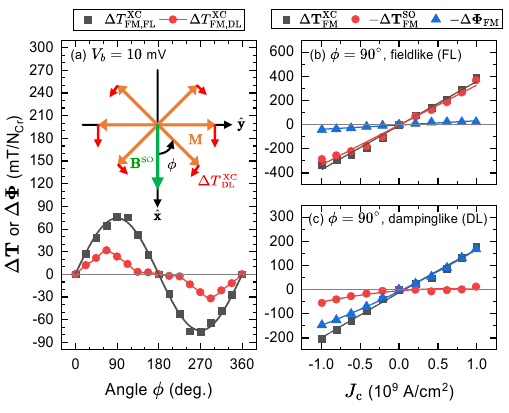}
	\caption{(a) Angular dependence of $\Delta \mathbf{T}^\mathrm{XC}_\mathrm{FM}$ in \ce{Cr3Te4} under an external bias of $V_b=10$ mV. The inset shows that $\mathbf{M}$ tends to align along the $x$ axis, due to the dampinglike torque $\Delta {\bf T}^\mathrm{XC}_\mathrm{FM,\mathrm{DL}}$ and precession motion from $\Delta {\bf T}^\mathrm{XC}_\mathrm{FM,FL}$, indicating a current-induced ${\bf B}^\mathrm{SO}$ along the $x$ direction. (b, c) Current dependence of $\Delta \mathbf{T}^\mathrm{XC}_\mathrm{FM}$, $-\Delta \mathbf{T}^\mathrm{SO}_\mathrm{FM}$, and $-\Delta \mathbf{\Phi}_\mathrm{FM}$ for the (b) fieldlike and (c) dampinglike components in \ce{Cr3Te4} with $\phi=90^\circ$ ($\mathbf{M}\parallel\hat{\mathbf{y}}$) corresponding to the maximum spin torques in (a). The solid curves represent linear or quadratic fits, with the linear term of $\Delta\mathbf{T}^\mathrm{XC}_\mathrm{FM}$ used to evaluate spin torque efficiency quantitatively.}
	\label{fig_cisot}
\end{figure}

Given that ${\bf B}^\mathrm{SO} \parallel \hat{\bf x}$ via Rashba-Edelstein charge-to-spin conversion, the spin torque vectors can be decomposed into the fieldlike (FL) and dampinglike (DL) components, aligned with $-\hat{\bf m} \times \hat{\bf x}$ and $-\hat{\bf m} \times \left( \hat{\bf m} \times \hat{\bf x} \right)$ directions, respectively.
Here, we consider the magnetization $\mathbf{M}=\hat{\bf m}\abs{\mathbf{M}}$, confined in the $x$-$y$ plane and rotated by an angle $\phi$, due to the IMA property discussed in Appendix~\ref{sec_sma}.
The current-induced quantities, $\Delta\mathbf{T}^\mathrm{XC}_\mathrm{FM}$, $\Delta\mathbf{T}^\mathrm{SO}_\mathrm{FM}$, and $\Delta\mathbf{\Phi}_\mathrm{FM}$, are defined by subtracting the zero-bias contribution, i.e., $\Delta\mathbf{T}^\mathrm{XC}_\mathrm{FM} = \mathbf{T}^\mathrm{XC}_\mathrm{FM}\left(V_b\right) - \mathbf{T}^\mathrm{XC}_\mathrm{FM}\left(V_b=0\right)$, and the same subtraction is applied to $\Delta\mathbf{\Phi}_\mathrm{FM}$ and ${\Delta\mathbf{T}}^\mathrm{SO}_\mathrm{FM}$.
Figure \ref{fig_cisot}(a) shows the angular dependency of $\Delta\mathbf{T}^\mathrm{XC}_\mathrm{FM}$ under an external bias of $V_b=10$ mV.
The sketch of the inset summarizes the quivers of the $\Delta {T}^\mathrm{XC}_\mathrm{FM,DL}$ vectors (red arrow).
This highlights the role of the spin-orbit field ${\bf B}^\mathrm{SO}$ (green arrow), which tend to align $\mathbf{M}$ along the $+x$ direction. The resulting motions include the dampinglike ($\Delta {T}^\mathrm{XC}_\mathrm{FM,DL}$) and precessional ($\Delta {T}^\mathrm{XC}_\mathrm{FM,FL}$) components.

To effectively distinguish the contributions of CISOT from the spin current injection and local spin induction, we display in Figs.~\ref{fig_cisot}(b)~and~\ref{fig_cisot}(c) the spin torques as a function of current density with $\phi=90^\circ$ ($\mathbf{M}\parallel\hat{\mathbf{y}}$), corresponding to the maximum value of Fig.~\ref{fig_cisot}(a).
For the fieldlike component, it is evident that $\Delta T_\mathrm{FM,FL}^\mathrm{XC} \approx -\Delta T_\mathrm{FM,FL}^\mathrm{SO}$, indicating the primarily contribution from the local spin induction within FM [Fig.~\ref{fig_mechanism}(c)].
From a previously studied 2D Rashba model \cite{Lee2015}, the current-induced spin density acts as an effective magnetic field, $\mathbf{B}^\mathrm{SO} \approx \abs{\bm \sigma}\hat{\mathbf{x}}$, generating a spin torque $\mathbf{T} \propto \mathbf{B}^\mathrm{SO} \times \mathbf{M}$, which primarily contributes to the fieldlike torque. This results in the nearly sinusoidal behavior observed in $\Delta{T}^\mathrm{XC}_{\mathrm{FM,FL}}$.
Also, the dominance of $\Delta T_\mathrm{FM,FL}^\mathrm{SO}$ can be attributed to the exchange splitting and the resulting evanescent state of the ferromagnetic layer \cite{Kim2017}.
%
On the other hand, the dampinglike torque, $\Delta T_\mathrm{FM,DL}^\mathrm{XC} \approx - \Delta\Phi_\mathrm{FM,DL}$, while smaller in magnitude than the fieldlike component, is contributed by spin current injection [Fig.~\ref{fig_mechanism}(b)].
Given that \ce{PtTe2} also exhibits small spin-splitting at the Fermi surface [Fig.~\ref{fig_struct_textures}(c)], additional spin current may be injected from NM \ce{PtTe2} into FM \ce{Cr3Te4}.

%
%
%

\subsection{Layer-resolved current-induced spin-orbit torque}
\label{sec_layer_cisot}

\begin{figure}[!b]
	\centering
	\includegraphics[scale=1]{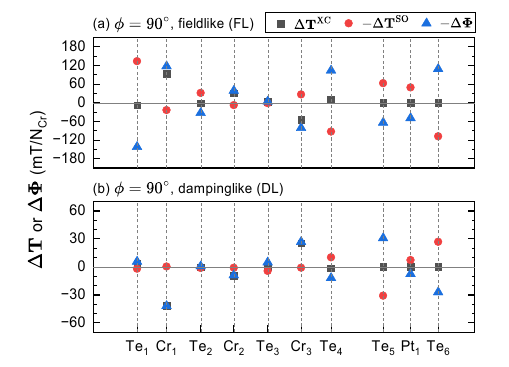}
	\caption{Layer-resolved (a) fieldlike and (b) dampinglike CISOT at $V_b = 10$ mV for the configuration $\phi = 90^\circ$ ($\mathbf{M} \parallel \hat{\mathbf{y}}$).}
	\label{fig_cisot_layer}
\end{figure}

In the previous section, we showed that $\Delta T_\mathrm{FM,FL}^\mathrm{XC} \approx -\Delta T_\mathrm{FM,FL}^\mathrm{SO}$ and $\Delta T_\mathrm{FM,DL}^\mathrm{XC} \approx -\Delta \Phi_\mathrm{FM,DL}$, highlighting that the FLT originates from local spin induction, whereas the DLT is governed by spin current injection. Because both the \ce{Cr3Te4} and \ce{PtTe2} layers contain Te atoms, which provide strong SOC, we now examine how each atomic layer contributes to the net CISOT.
Figure \ref{fig_cisot_layer} presents the layer-resolved CISOT at $V_b = 10$ mV for the configuration $\phi = 90^\circ$ ($\mathbf{M} \parallel \hat{\mathbf{y}}$). As expected, the exchange spin torque $\Delta {\bf T}^{\mathrm{XC}}$ acts primarily on the magnetic Cr atoms, and both the fieldlike [Fig.~\ref{fig_cisot_layer}(a)] and dampinglike [Fig.~\ref{fig_cisot_layer}(b)] components of $\Delta {\bf T}^{\mathrm{XC}}$ are concentrated on the interfacial Cr sites (Cr\textsubscript{1} and Cr\textsubscript{3}). This clearly reflects a surface-localized torque, consistent with the interfacial nature of ISOC.

From the net FLT shown in Fig.~\ref{fig_cisot}(b), we know that the dominant torque within \ce{Cr3Te4} arises from the spin-orbit torque, i.e., $\Delta T_\mathrm{FM,FL}^\mathrm{XC} \approx -\Delta T_\mathrm{FM,FL}^\mathrm{SO}$. In Fig.~\ref{fig_cisot_layer}(a), the layer-resolved analysis reveals that the boundary Te atoms (Te\textsubscript{1} and Te\textsubscript{4}) generate particularly strong $\Delta T^\mathrm{SO}_\mathrm{Te,FL}$, revealing the crucial role of Te-driven SOC. Because these Te atoms are essentially nonmagnetic, their exchange torque $\Delta T^\mathrm{XC}_\mathrm{Te,FL}$ and $\Delta T^\mathrm{XC}_\mathrm{Te,DL}$ are nearly zero. Instead, the influence of Te on the magnetic Cr atoms is transmitted through spin-current flow, quantified by $\Delta {\bf \Phi}$.
A nonzero $\Delta {\bf \Phi}$ represents the presence of a spin source or sink \cite{Huang2023b}: It indicates where spin angular momentum is injected or absorbed within the layer.
For example, the opposite signs of $\Delta\Phi_\mathrm{FL}$ for Te\textsubscript{1} and Cr\textsubscript{1} (also Cr\textsubscript{3} and Te\textsubscript{4}) indicates a spin flow between them.
Importantly, the total spin current flow within the \ce{Cr3Te4} region satisfies $\sum_{i \in \mathrm{FM}}{\Delta \Phi_{i,\mathrm{FL}}} \approx 0$ in Fig.~\ref{fig_cisot}(b), showing that spin angular momentum originating from the Te-layer SOC is redistributed internally within the \ce{Cr3Te4} layer. There is no net spin inflow or outflow between \ce{Cr3Te4} and \ce{PtTe2}, consistent with the discussion in Sec.~\ref{sec_cisot}, where the local spin induction within the FM layer via Te atoms dominates the FLT.

For DLT shown in Fig.~\ref{fig_cisot_layer}(b), the spin-orbit torque $\Delta T^\mathrm{SO}_\mathrm{Te,DL}$  arises predominantly from the \ce{PtTe2} layer. The nonzero $\Delta \Phi_\mathrm{DL}$ in both the \ce{Cr3Te4} and \ce{PtTe2} layers indicates a spin current flows across the interface, which in turn triggers the dampinglike exchange torque on \ce{Cr3Te4}, consistent with the relation $\Delta T_\mathrm{FM,DL}^\mathrm{XC} \approx -\Delta \Phi_\mathrm{FM,DL}$, where the spin current injection from \ce{PtTe2} contributes the DLT.


In summary, our layer-resolved analysis demonstrates the crucial role of Te-driven SOC in generating CISOT.
For FLT on \ce{Cr3Te4}, the boundary Te atoms produce a strong local spin-orbit torque $\Delta T^\mathrm{SO}_\mathrm{Te,FL}$, which is transferred to the magnetic Cr sites through a nonzero divergence of spin current $\Delta \Phi_\mathrm{FL}$.
For DLT, the Te atoms in the \ce{PtTe2} layer similarly generate a significant  $\Delta T^\mathrm{SO}_\mathrm{Te,DL}$, and this torque is injected into \ce{Cr3Te4} via the interlayer spin current reflected in $\Delta \Phi_\mathrm{DL}$. This spin-current inflow leads to a net spin accumulation on the magnetic Cr atoms and consequently produces the DLT.
Overall, Te atoms act as the primary SOC sources, while \ce{Cr3Te4} receives the torque mainly through spin-current transfer rather than local SOC in the Cr atoms.

\subsection{Equilibrium and nonequilibrium contributions to spin-orbit torques}
\label{sec_eq_neq_cisot}

\begin{figure}
	\centering
	\includegraphics[scale=1]{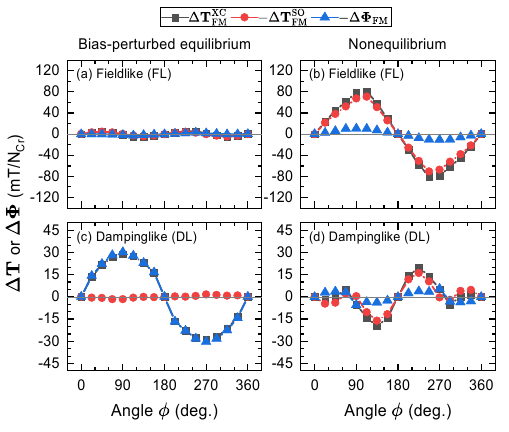}
	\caption{Decomposed angular dependence of (a,b) fieldlike torque (FLT) and (c,d) dampinglike torque (DLT) in \ce{Cr3Te4} under an external bias of $V_b=10$ mV. The decomposition includes (a,c) the bias-perturbed equilibrium and (b,d) the nonequilibrium contributions.}
	\label{fig_cisot_eq_neq}
\end{figure}

Within the Kubo linear-response formalism, Freimuth \textit{et al}. \cite{Freimuth2014} demonstrated that CISOT can be decomposed into an even ($T^\mathrm{even}$) and an odd ($T^\mathrm{odd}$) component with respect to magnetization reversal. Specifically, $T^\mathrm{even}$ arises from the intrinsic Berry curvature of the occupied states (the Fermi sea), whereas $T^\mathrm{odd}$ is driven entirely by Fermi surface terms.
To draw a direct mapping between our NEGF calculations and this Kubo-based framework, we partition the energy integration of the spin torque into an occupied-state contribution and a bias-window contribution: $\mathbf{T} = \int_{-\infty}^{\mu_1} \mathbf{T}(E) dE + \int_{\mu_1}^{\mu_2} \mathbf{T}(E) dE$, where $\mu_1$ and $\mu_2$ (with $\mu_2 > \mu_1$) are the chemical potentials of the two electrodes.
This partitioning enables us to separate the total CISOT into a bias-perturbed equilibrium component ($\Delta \mathbf{T}^{\mathrm{eq}}$) and a purely nonequilibrium component ($\mathbf{T}^{\mathrm{neq}}$):
\begin{equation}
	\Delta \mathbf{T} = \mathbf{T}(V_b) - \mathbf{T}(V_b=0) = \Delta \mathbf{T}^{\mathrm{eq}} + \mathbf{T}^{\mathrm{neq}}
\end{equation}
with the isolated terms defined as
\begin{align}
	\Delta \mathbf{T}^{\mathrm{eq}} &= \int_{-\infty}^{\mu_1} \mathbf{T}(E, V_b) dE - \int_{-\infty}^{0} \mathbf{T}(E, V_b=0) dE
	\\
	\mathbf{T}^{\mathrm{neq}} &= \int_{\mu_1}^{\mu_2} \mathbf{T}(E, V_b) dE ,
\end{align}
which physically map to the Fermi sea and Fermi surface contributions, respectively.

Figure~\ref{fig_cisot_eq_neq} presents the bias-perturbed equilibrium and nonequilibrium components of the angular-dependent spin torques under an external bias of $V_b=10$ mV.
Given that $\Delta {\bf T}^\mathrm{XC}_\mathrm{FM}$ represents the total spin torque acting on the magnetic \ce{Cr3Te4} layer, Figs.~\ref{fig_cisot_eq_neq}(b) and \ref{fig_cisot_eq_neq}(c) demonstrate two distinct mechanisms: The FLT is predominantly driven by the nonequilibrium component, whereas the DLT originates mostly from the bias-perturbed equilibrium component with only a minor nonequilibrium contribution to the DLT shown in Fig.~\ref{fig_cisot_eq_neq}(d).
These results are in agreement with the Kubo-response framework of Ref.~\cite{Freimuth2014}, which established that FLT and DLT arise from Fermi surface and Fermi sea contributions, respectively, consistent with our decomposition into the bias-window and occupied-state integrals.
Furthermore, by partitioning $\Delta {\bf T}^\mathrm{XC}_\mathrm{FM}$ into a local spin-orbit torque ($\Delta {\bf T}^\mathrm{SO}_\mathrm{FM}$) and a spin current accumulation ($\Delta {\bf \Phi}_\mathrm{FM}$), we find that the bias-perturbed equilibrium DLT is dominated by $\Delta {\bf \Phi}_\mathrm{FM}$, while the nonequilibrium FLT is driven primarily by the local $\Delta {\bf T}^\mathrm{SO}_\mathrm{FM}$.

Based on these observations, we can definitively assign the  physical origins of the torques generated by the \ce{PtTe2} and \ce{Cr3Te4} layers.
For \ce{PtTe2}, it provides a spin current source originated from the intrinsic spin Hall effect (SHE) and generates a net spin current accumulation $\Delta {\bf \Phi}_\mathrm{FM}$ in \ce{Cr3Te4}. The intrinsic SHE is a topological phenomenon mathematically governed by the mixed $\mathbf{k}$-$\hat{\bf M}$ Berry curvature of the occupied states \cite{Freimuth2014}. This matches our finding in Fig.~\ref{fig_cisot_eq_neq}(c) that the DLT is governed by the bias-perturbed equilibrium (Fermi sea) contribution. Moreover, as shown in Fig.~\ref{fig_cisot_layer}, the Te\textsubscript{5} and Te\textsubscript{6} atoms in the \ce{PtTe2} layer exhibit opposite signs for both $\Delta \Phi_\mathrm{FL}$ and $\Delta \Phi_\mathrm{DL}$.
For a magnetization $\mathbf{M} \parallel \hat{\bf y}$, the DLT and FLT vectors point along the $x$ and $z$ axes, respectively. This indicates a clear spatial separation of $\pm x$ and $\pm z$ spin current sources across the two Te layers, namely, a signature of hidden spin polarization. Notably, while both spin components are present, only the equilibrium-driven $x$-polarized spin current flux effectively drives the DLT on the adjacent magnetic layer.

Conversely, the torque originating within the \ce{Cr3Te4} layer is driven by the Rashba-Edelstein effect (REE), as discussed in Sec.~\ref{sec_cisot}. The interfacial inversion asymmetry induces Rashba spin-splitting near the Fermi energy. When a longitudinal charge current is applied along the $y$ direction, this splitting yields a nonequilibrium $x$-polarized spin accumulation (the Edelstein effect), which acts as an effective internal magnetic field, $\mathbf{B}^\mathrm{SO}$. The nonequilibrium spin torques shown in Fig.~\ref{fig_cisot_eq_neq} clearly reflect this kinetic mechanism, which appears primarily as the local spin induction $\Delta {\bf T}^\mathrm{SO}_\mathrm{FM}$ generated directly within the \ce{Cr3Te4} layer.

\subsection{Current-induced spin density}

\begin{figure}
	\centering
	\includegraphics[scale=1]{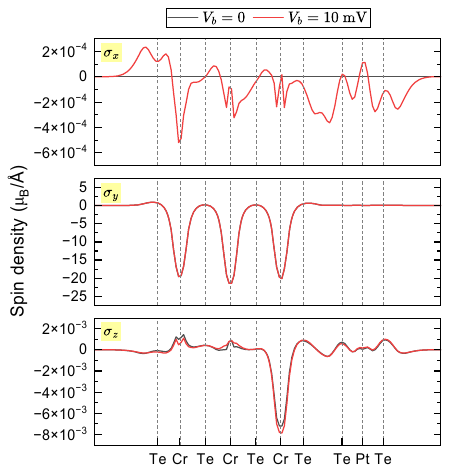}
	\caption{Current-induced spin density along the real-space $z$ direction at $V_b = 0$ and $V_b = 10$ mV. The magnetization direction of the \ce{Cr3Te4} layer is $\mathbf{M}\parallel\hat{\bf y}$.}
	\label{fig_cispin}
\end{figure}

In this section, we quantitatively examine the current-induced spin density and its relation to the CISOT. Figure~\ref{fig_cispin} shows the spin density along the real-space $z$ direction at equilibrium ($V_b = 0$) and under a small applied bias ($V_b = 10$ mV). The magnetization of the \ce{Cr3Te4} layer is fixed at $\phi=90^\circ$ ($\mathbf{M}\parallel\hat{\bf y}$), which corresponds to the maximum CISOT observed in Fig.~\ref{fig_cisot}(a).
At $V_b = 0$ V, the nonzero spin density primarily reflects the intrinsic magnetization of \ce{Cr3Te4} on $\sigma_y$. A small but finite $\sigma_z$ component remains even when $\mathbf{M}\parallel\hat{\mathbf{y}}$, which arises from the fully self-consistent NEGF-DFT calculation and is not constrained to be zero.
Under $V_b = 10$ mV, a clear current-induced $x$ component of the spin density emerges. This component corresponds to an effective spin-orbit field ${\bf B}^\mathrm{SO}\parallel\hat{\mathbf{x}}$, which drives the CISOT observed in Fig.~\ref{fig_cisot}. As discussed in Sec.~\ref{sec_spin_textures}, for charge transport along the $y$ direction, the spin texture shows that the $\sigma_x$ component possesses inversion symmetry with respect to $k_y$, contributing to the Rashba-Edelstein effect that generates ${\bf B}^\mathrm{SO} \parallel \hat{\bf x}$.

\subsection{Macrospin dynamics}

%
So far, we have discovered the pivotal role of the local spin induction ($ \Delta T_\mathrm{FM,FL}^\mathrm{SO}$) in the FLT and the spin current accumulation ($\Delta\Phi_\mathrm{FM,DL}$) for the DLT.
We next demonstrate magnetization switching using a macrospin simulation approach. The dynamics of magnetization are numerically simulated by solving the Landau-Lifshitz-Gilbert (LLG) equation with additional spin torque terms \cite{Sun2000,Xiao2005}:
\begin{equation}\label{eq_llg}
	\begin{aligned}
		\frac{1}{\gamma} \frac{d\hat{\bf m}}{dt} = &-\hat{\bf m} \times \mu_0\mathbf{H}_\mathrm{eff} + \frac{\alpha}{\gamma}\hat{\bf m} \times \frac{d\hat{\bf m}}{dt}
		\\
		&- \hat{\bf m} \times \left(\hat{\bf m} \times \mu_0\tau_\mathrm{DL}J_\mathrm{c}\hat{\bf x}\right) - \hat{\bf m} \times \mu_0\tau_\mathrm{FL}J_\mathrm{c}\hat{\bf x},
	\end{aligned}
\end{equation}
where $\hat{\bf m}=\mathbf{M}/\abs{\mathbf{M}}$ is the unit vector of the FM magnetization, $\gamma$ is the gyromagnetic ratio, $\mathbf{H}_\mathrm{eff}$ is the effective magnetic field, $\alpha$ is the damping constant, the charge current density $J_\mathrm{c}$ is applied along the $y$ axis, and $\hat{\bf x}$ represents the direction of current-induced ${\bf B}^\mathrm{SO}$, i.e., the green arrow in the inset of Fig.~\ref{fig_cisot}(a).

The linear term of linear/quadratical fitting of $\Delta\mathbf{T}^\mathrm{XC}_\mathrm{FM}$, i.e., the solid curves in Figs.~\ref{fig_cisot}(b)~and~\ref{fig_cisot}(c), allows us to quantitatively estimate the spin torque efficiency at the fieldlike ($\tau_\mathrm{FL}$) and dampinglike ($\tau_\mathrm{DL}$) direction, which are $\mu_0\tau_\mathrm{FL}=3.67\times{10}^{-7}\ \mathrm{mT/(A\ cm^{-2}}$) and $\mu_0\tau_\mathrm{DL}=1.84\times{10}^{-7}\ \mathrm{mT/(A\ cm^{-2})}$. Similar to the spin Hall angle, the dimensionless spin torque efficiency \cite{Liu2012}, defined as $\eta \equiv (2e/\hbar) M_\mathrm{s} t_\mathrm{FM} \mu_0 \tau$, is $\eta_\mathrm{FL} = 0.076$ and $\eta_\mathrm{DL} = 0.038$.
As discussed in Ref.~\cite{Tian2021}, the spin torque efficiency in vdW-based materials typically ranges from 0.005 to 0.05, which is lower compared to those of heavy metal-based materials with efficiency of approximately 0.05 to 0.5. This is reasonable because the Rashba electric field is localized near the interface, unlike the bulk spin Hall effect, in which SOC influences the entire NM region.

Based on DFT calculations in Table~\ref{tab_struct_params}, the FM layer thickness is $t_\mathrm{FM}=0.97\ \mathrm{nm}$, and the saturation magnetization $M_\mathrm{s}=\abs{\mathbf{M}}$ is $708 \times{10}^3\ \mathrm{A/m}$.
The easy-plane anisotropy is given by $K_\mathrm{plane} = K_\mathrm{demg} + K_\mathrm{surf} / t_\mathrm{FM}$, where the demagnetization energy for a magnetic thin film can be estimated by $K_\mathrm{demg} = -(1/2) \mu_0 M_\mathrm{s}^2$, and the surface anisotropy energy, $K_\mathrm{surf} = -0.56\ \mathrm{mJ/m^2}$, is determined from the equilibrium magnetic anisotropy energy (see Appendix~\ref{sec_sma}). The negative sign of $K_\mathrm{surf}$ indicates that \ce{Cr3Te4} has a hard axis along the $z$ direction, confirming as an IMA system. The corresponding hard-axis effective field along the $z$ direction is represented as $\mu_0 H_\mathrm{p} =  2 K_\mathrm{plane} / M_\mathrm{s} = -2.51$ T.
Here we choose $\alpha=0.01$ and a small uniaxial anisotropy, $K_\mathrm{uni} = (1/2) \mu_0 M_\mathrm{s} H_\mathrm{k}$ to represent an easy axis aligned along the $x$ axis with the field magnitude $\mu_0 H_\mathrm{k}=150 \times {10}^{-4}\ \mathrm{T}$, adjustable based on the aspect ratio of the magnetic thin film \cite{Xiao2005}.
To align with comparison to real memory devices, we choose the cell area to be $A=39100\ \mathrm{nm^2}$ (or $\sqrt A\approx198\ \mathrm{nm}$ roughly showing the dimension of device). Thus, at room temperature ($T=300~\mathrm{K}$), the thermal stability factor is given by $\Delta = \mu_0 M_\mathrm{s} H_\mathrm{k} A t_\mathrm{FM}/(2k_\mathrm{B} T) \approx 48$, where a value in the range of 40 to 50 corresponds to a data retention time of ten years \cite{Sun2000,Timopheev2015}.

\begin{figure}
	\centering
	\includegraphics[scale=1]{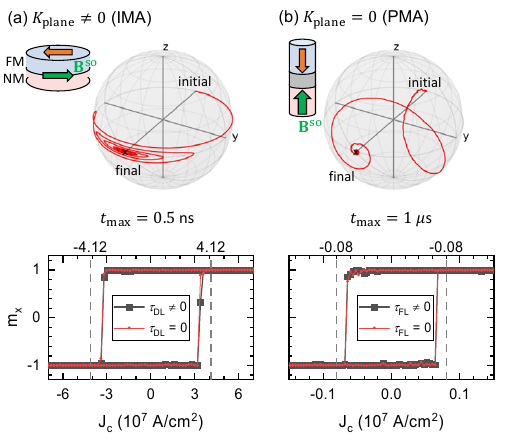}
	\caption{(a) The IMA system with and without dampinglike spin torque efficiency ($\tau_\mathrm{DL}$). (b) The PMA system with and without fieldlike spin torque efficiency ($\tau_\mathrm{FL}$). [Top] The precession motion of the magnetization under a current density $J_\mathrm{c}=5\times{10}^7\ \mathrm{A/cm^2}$ at 300K. [Bottom] Magnetization in the $x$ direction as a function of the applied current ($M$-$J_\mathrm{c}$ loops). The dashed lines of $M$-$J_\mathrm{c}$ loops represent the estimated $J_{\mathrm{sw}}$ from Eqs. \eqref{eq_switching_2} and \eqref{eq_sjpsoq} for (a) IMA and (b) PMA cases, respectively.}
	\label{fig_llg}
\end{figure}

We present in Fig.~\ref{fig_llg}(a) the magnetization switches from $-x$ to $+x$ under a current density of $J_\mathrm{c}=5\times{10}^7\ \mathrm{A/cm^2}$ at 300 K, where the strong easy-plane anisotropy $K_\mathrm{plane}$ confines the magnetization to precess nearly in the $x$-$y$ plane.
The corresponding hysteresis loops of magnetization versus current ($M$-$J_\mathrm{c}$ loops) are obtained using a fast switching scheme with the duration time of $t_\mathrm{max}(\mathrm{IMA}) = 0.5\ \mathrm{ns}$ for each current density.
With ($\tau_\mathrm{DL} \ne 0$) and without ($\tau_\mathrm{DL} = 0$) the DLT efficiency, the nearly unchanged switching current density, $J_\mathrm{sw} \approx 3\times{10}^7\ \mathrm{A/cm^2}$, reveals a fact that the FLT dominates the current-driven magnetization switching of IMA.
This is very different from the most magnetic bilayers, where the DLT is typically considered the primary determinant of the switching current, as it controls the precession angle between stable points along the easy axis \cite{Sun2000,Liu2012,Timopheev2015,Fukami2016}.
As noted in Ref.~\cite{Sun2000}, under strong easy-plane anisotropy conditions, the switching current can become dependent on an external magnetic field, even though the damping constant reduces the influence of the external field.
Given that $\tau_\mathrm{FL}$ in Eq.~\eqref{eq_llg} takes the same form as the effective field, the magnetic switching condition must be adjusted accordingly.

When the easy-plane anisotropy, $K_\mathrm{plane}$, is strong enough to confine the magnetization's precession within the $x$-$y$ plane, the steady state condition $dm_z/dt \approx 0$ leads to the threshold criterion (see Appendix~\ref{sec_derivation}):
\begin{equation}\label{eq_switching}
	H_\mathrm{k} + \tau_\mathrm{FL}J_\mathrm{c} - \alpha \tau_\mathrm{DL}J_\mathrm{c} = 0 .
\end{equation}
It is evident that the FLT ($\tau_\mathrm{FL}$) primarily governs the critical switching field, since the weighted DLT ($\alpha\tau_\mathrm{DL}$) by the damping constant $\alpha$ (typically smaller than 0.01) is considerably weaker.
%
Rewriting Eq.~\eqref{eq_switching}, the switching current for strong-IMA systems can be expressed as
\begin{equation}\label{eq_switching_2}
	J_{\mathrm{sw}}(\text{strong--IMA})=\frac{2e}{\hbar}\frac{\mu _0M_{\mathrm{s}}t_{\mathrm{FM}}}{\eta _{\mathrm{FL}}{ -\alpha \eta _\mathrm{DL}}} H_\mathrm{k} .
\end{equation}
This model predicts a switching current of $J_\mathrm{sw} = 4.12 \times 10^7\ \mathrm{A/cm^2}$, as marked by the dashed lines in the hysteresis loops of Fig.~\ref{fig_llg}(a). Furthermore, it highlights that FLT dominates the switching current via the ISOC-induced local spin induction mechanism, in all-vdW \ce{Cr3Te4}/\ce{PtTe2} with strong-IMA configuration.

Note that for weak-IMA systems, where $K_\text{plane}$ is not strong enough to confine the trajectory within the $x$-$y$ plane, the switching condition can be derived using energy stability analysis \cite{Sun2000, Bazaliy2004}, which is expressed as
\begin{equation}
	\alpha \left( H_{\mathrm{k}}+\frac{1}{2}|H_{\mathrm{p}}|+\tau _{\mathrm{FL}}J_{\mathrm{c}} \right) +\tau _{\mathrm{DL}}J_{\mathrm{c}}=0 ,
\end{equation}
leading to the expression for the switching current:
\begin{equation}\label{eq_sjpsoq}
	J_{\mathrm{sw}}(\text{weak--IMA})=\frac{2e}{\hbar}\frac{\mu _0M_{\mathrm{s}}t_{\mathrm{FM}}}{\eta _{\mathrm{FL}}+\frac{1}{\alpha}\eta _{\mathrm{DL}}}\left( H_{\mathrm{k}}+\frac{1}{2}|H_{\mathrm{p}}| \right) ,
\end{equation}
where the absolute value is added to $H_{\mathrm{p}}$ since the negative value represents a hard axis in our derivations. Note that Eq.~\eqref{eq_sjpsoq} can be used to describe PMA systems by removing the easy-plane anisotropy ($H_\text{p}=0$). We also provide an alternative derivation for PMA switching current in Appendix~\ref{sec_derivation}.

For comparison with PMA systems, such as MgO/Fe-based heterojunctions \cite{Sun2000,Timopheev2015}, Fig.~\ref{fig_llg}(b) illustrates the magnetic trajectory and $M$-$J_\mathrm{c}$ loops over an extended simulation duration of $t_\mathrm{max}(\mathrm{PMA}) = 1\ \mathrm{\mu s}$.
All parameters are identical except for setting $K_\mathrm{plane}=0$ and $H_\text{p}=0$ to represent PMA. Using Eq.~\eqref{eq_sjpsoq}, this gives a value of $J_{\mathrm{sw}} = 0.08\times 10^7$ $\mathrm{A}/\mathrm{cm}^2$, as marked by the dashed lines in the hysteresis loops of Fig.~\ref{fig_llg}(b).
Compared to Eq.~\eqref{eq_switching_2}, we see that an additional $H_\text{p}/2$ term appears to account for the easy-plane anisotropy. However, in the regime of strong-IMA, Eq.~\eqref{eq_switching_2} is not a function of $H_\text{p}$, keeping $J_\text{sw}$ a constant.
Furthermore, the $1/\alpha$ term in the denominator enhances the contribution of $\eta _{\mathrm{DL}}$, causing the DLT in PMA or weak-IMA systems to dominate the switching current. This dominance explains the nearly identical $M$-$J_\mathrm{c}$ loops observed with and without FLT efficiency, $\tau_\mathrm{FL}$, in Fig.~\ref{fig_llg}(b).


\begin{figure}
	\centering
	\includegraphics[scale=1]{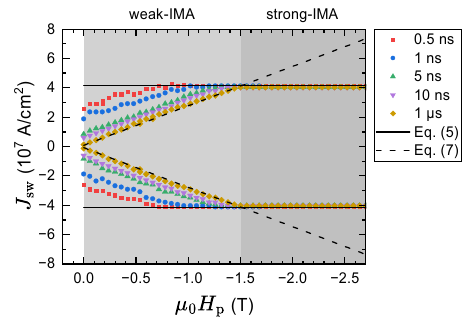}
	\caption{Switching current $J_\text{sw}$ as a function of $\mu_0 H_\text{p}$ for different pulse durations $t_\text{max}$ at zero temperature ($T=0$). Positive $J_\text{sw}$ corresponds to magnetization switching from $-x$ to $+x$, while negative $J_\text{sw}$ corresponds to switching from $+x$ to $-x$. A maximum in $J_\text{sw}$ appears with increasing $|\mu_0 H_\text{p}|$, marking the turning point that separates the weak-IMA and strong-IMA regimes.}
	\label{fig_Hp_phase}
\end{figure}

To clarify the regimes under which Eq.~\eqref{eq_switching_2} or Eq.~\eqref{eq_sjpsoq} should be applied to IMA systems, we plot in Fig.~\ref{fig_Hp_phase} the switching current as a function of $H_\text{p}$ for different duration time $t_\text{max}$, ranging from 0.5 ns to 1 $\mu$s.  To better compare with the analytical expressions for $J_\text{sw}$, the simulations are performed at $T=0$.
In all cases, the switching current exhibits a maximum as $H_\text{p}$ increases, regardless of its absolute magnitude.
In the weak-IMA regime and for sufficiently long durations ($t_\text{max} \gtrsim 5$ ns), $J_\text{sw}$ is well described by Eq.~\eqref{eq_sjpsoq}.
Beyond a turning point, however, the switching current must instead be evaluated using Eq.~\eqref{eq_switching_2}.
The turning point is determined by solving Eq.~\eqref{eq_switching_2} and Eq.~\eqref{eq_sjpsoq}, giving
\begin{equation}
	|\tilde{H}_{\mathrm{p}}|=2H_{\mathrm{k}} \frac{\eta _{\mathrm{DL}}\left( \frac{1}{\alpha}+\alpha \right)}{\left( \eta _{\mathrm{FL}}-\alpha \eta _{\mathrm{DL}} \right)} .
\end{equation}
In the typical limit $\alpha \ll 1$, this simplifies to
\begin{equation}
	|\tilde{H}_{\mathrm{p}}| =2H_{\mathrm{k}} \frac{1}{\alpha}\frac{\eta _{\mathrm{DL}}}{\eta _{\mathrm{FL}}} ,
\end{equation}
which gives $|\tilde{H}_{\mathrm{p}}| \approx 1.5$ T in our case. Thus, the turning point scales approximately as $1/\alpha$, with $\alpha$ usually lying in the range $0.001$--$0.1$.
In many practical situations including PMA and weak-IMA systems, Eq.~\eqref{eq_sjpsoq} is sufficient to accurately describe the switching current observed in experiments.
However, in thin films with strong surface-induced IMA, the easy-plane anisotropy energy $K_\mathrm{plane} = K_\mathrm{demg} + K_\mathrm{surf}/t_\mathrm{FM}$ can become large enough to push the system beyond the turning point. In this regime, the switching current is instead well described by Eq.~\eqref{eq_switching_2}, and the FLT becomes dominance to the strong-IMA switching.
For our bilayer, $\mu_0 H_\text{p} = \mu_0 H_\text{demg} + \mu_0 H_\text{surf} = -2.51 $ T, where $\mu_0 H_\text{demg} = -\mu_0 M_\text{s} = -0.89$ T and $\mu_0 H_\text{surf} = 2K_\text{surf}/(M_\text{s} t_\text{FM}) = -1.62$ T. We see that the large surface-induced IMA $K_\text{surf}$ push the system into the strong-IMA regime, which goes beyond the turning point.


Through the study of LLG magnetization switching in both strong-IMA and PMA systems, we conclude that switching current is primarily governed by the FLT and DLT, respectively.
Strong-IMA switching, driven by FLT via ISOC-induced local spin induction, requires higher switching current density but achieves significantly faster switching compared to PMA. This makes IMA switching advantageous for high-speed writing applications.
Conversely, PMA or weak-IMA switching, dominated by DLT from spin injection mechanism, demands lower switching current, making it more suitable for energy-efficient memory technologies.
Given the natural IMA of \ce{Cr3Te4}, attributed to proximity effects from \ce{PtTe2}, the \ce{Cr3Te4}/\ce{PtTe2} heterobilayer emerges as a promising platform for SOT-based MRAM devices utilizing all-2D materials.

%

\section{Conclusions}

In conclusion, we have presented techniques for differentiating between spin current injection and local spin induction mechanisms of magnetic hetrobilayers using spin current accumulation and spin-orbit torque analyses. Our study of the vdW heterobilayer, \ce{Cr3Te4}/\ce{PtTe2}, reveals that the ferromagnetic \ce{Cr3Te4} exhibits significant Rashba spin-splitting due to the interfacial effect induced from \ce{PtTe2}. When an in-plane charge current is applied, the device shows a larger fieldlike torque than dampinglike torque. We identify that the local spin induction, resulting from the Rashba spin-orbit coupling in \ce{Cr3Te4}, contributes to this fieldlike torque.
This finding provides a link between the generation of fieldlike torque and research in material databases.
Moreover, we show that the fieldlike torque plays a significant role in current-driven magnetization switching under strong-IMA conditions. Unlike devices with PMA, which require an external magnetic field for deterministic switching, the IMA-based devices, in conjunction with fieldlike torque, suggests a promising pathway for designing field-free SOT-based memory devices.

\section*{Acknowledgment}

Y.-H.T. and B.-H.H. acknowledge the National Science and Technology Council, Taiwan (NSTC 108-2628-M-008-004-MY3 / 111-2112-M-008-025 / 112-2112-M-008-036), the National Center for Theoretical Sciences, and the National Center for High-performance Computing for providing computational and storage resources. H.G. thanks NSERC of Canada for financial support. We gratefully acknowledge  Nanoacademic Technologies Inc. of Canada for providing the NEGF-DFT quantum transport package NanoDCAL used in this work.
The authors would thank to Hsin Lin and Sheng-Chieh Huang for valuable discussions and feedback on the manuscript. 



\appendix

\section{Note on the NEGF-DFT calculations using LDA functional}
\label{sec_note_lda}

\begin{figure}
	\centering
	\includegraphics[scale=1]{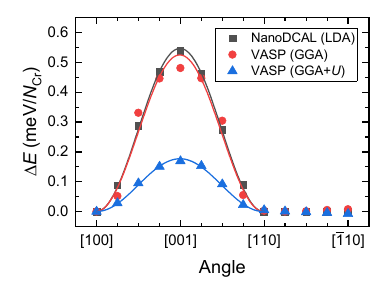}
	\caption{Magnetic anisotropy energy (MAE, $\Delta E$) for various magnetization angles in the \ce{Cr3Te4}/\ce{PtTe2} bilayer. The solid curves are fitted by using $\sin^2\theta$ function. For VASP calculations with GGA+$U$, $U=4.0$ eV for Cr atoms is included.}
	\label{fig_sma}
\end{figure}

Due to the lack of GGA-based basis sets for the Te atom, we performed our NEGF-DFT calculations using the LDA XC functional. To validate these calculations, we employed the VASP package with the GGA-based PBE XC functional to compute equilibrium electronic properties, including the projected band structures (Fig.~\ref{fig_bands_projected}), spin textures at the Fermi energy (Fig.~\ref{fig_spin_textures}) and magnetic anisotropy energy (MAE, Fig.~\ref{fig_sma}).
The comparisons between the \textsc{NanoDCAL}-LDA and \textsc{VASP}-GGA results without Hubbard $U$ show qualitative agreement, supporting the reliability of our calculations.

\section{Detailed band structures and spin textures}
\label{sec_detail_bands}

\begin{figure*}
	\centering
	\includegraphics[scale=1]{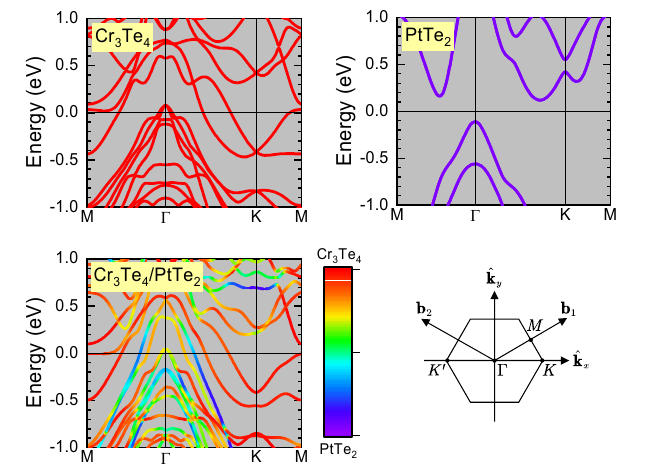}
	\caption{Band structures of the \ce{Cr3Te4} monolayer, \ce{PtTe2} monolayer, and \ce{Cr3Te4}/\ce{PtTe2} heterobilayer. The magnetization direction of \ce{Cr3Te4} is aligned along the $x$ axis, $\mathbf{M}\parallel\hat{\bf x}$. The color shows the proportion between \ce{Cr3Te4} and \ce{PtTe2}.}
	\label{fig_bands_general}
\end{figure*}

In Fig.~\ref{fig_bands_general}, we present the band structures of the \ce{Cr3Te4} monolayer, \ce{PtTe2} monolayer, and \ce{Cr3Te4}/\ce{PtTe2} heterobilayer. The magnetization direction of \ce{Cr3Te4} is aligned along the $y$ axis ($\mathbf{M}\parallel\hat{\bf y}$), with SOC included. The \ce{PtTe2} monolayer exhibits an indirect band gap of approximately 0.23 eV. In contrast, both the \ce{Cr3Te4} monolayer and the \ce{Cr3Te4}/\ce{PtTe2} heterobilayer are gapless, indicating their metallic properties. Near the $\Gamma$-point, the green and light blue colors indicate hybridization between the \ce{Cr3Te4} and \ce{PtTe2} layers.

\begin{figure*}[b]
	\centering
	\includegraphics[scale=1]{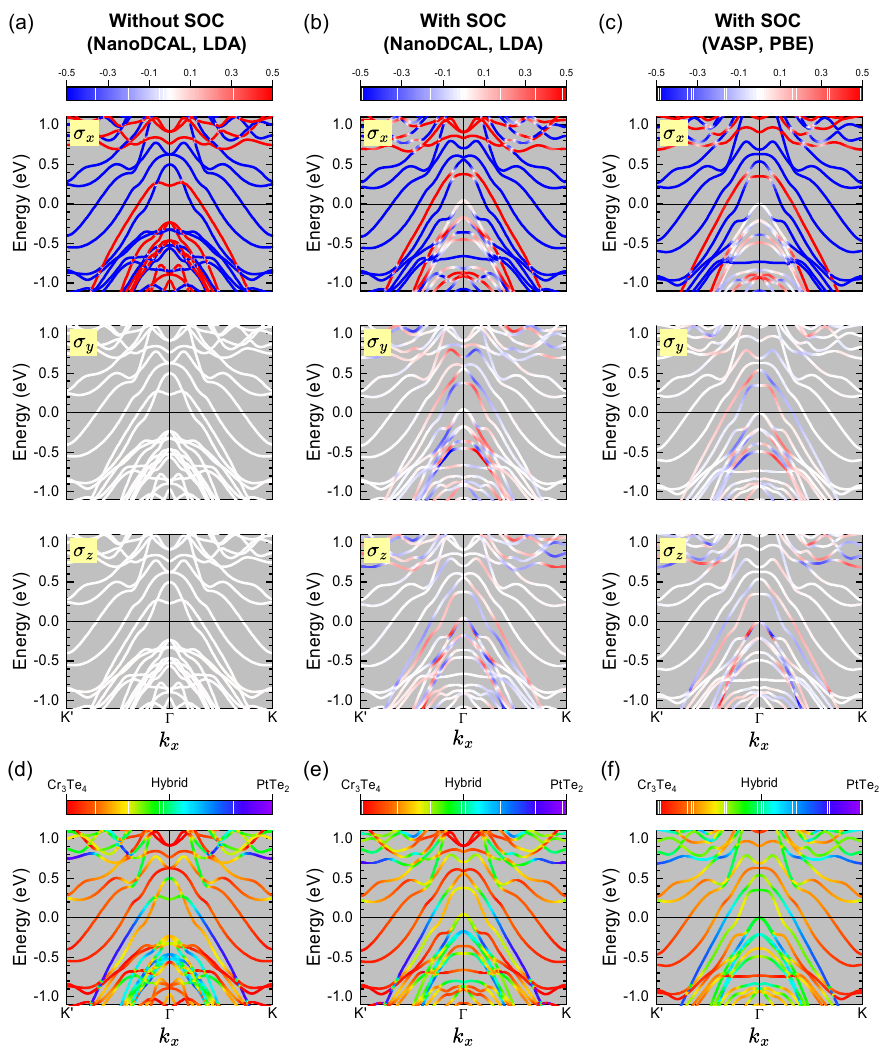}
	\caption{(a), (b), (c) show the band structures projected onto the Pauli matrices ($\sigma_x$, $\sigma_y$, and $\sigma_z$), along the $\hat{\bf k}_x$ direction. (d), (e), (f) depict the band structures projected onto the proportion between \ce{Cr3Te4} and \ce{PtTe2}. (a) and (d) are without SOC, while (b), (c) and (e), (f) include SOC. (c) and (f) provide the comparison using VASP-PBE calculation. The magnetization direction of \ce{Cr3Te4} is aligned along the $x$ axis, $\mathbf{M}\parallel\hat{\bf x}$.}
	\label{fig_bands_projected}
\end{figure*}

\begin{figure*}[b]
	\centering
	\includegraphics[scale=1]{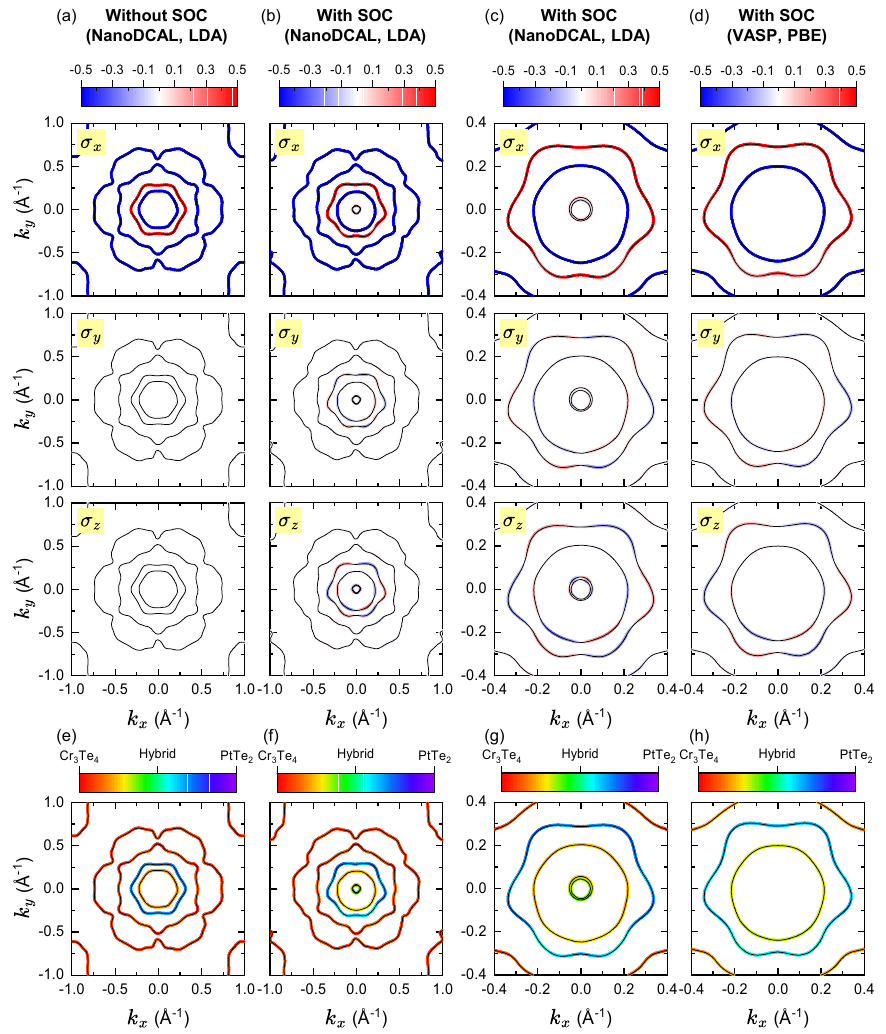}
	\caption{Same as Fig.~\ref{fig_bands_projected}, but at the Fermi energy, representing the spin textures. (c) provides an enlarged view of the regions shown in (b) near the $\Gamma$-point. (d) provides the comparison using VASP-PBE calculation.}
	\label{fig_spin_textures}
\end{figure*}

Figures~\ref{fig_bands_projected}(a)--\ref{fig_bands_projected}(c) display the band structures projected onto the Pauli matrices ($\sigma_x$, $\sigma_y$, and $\sigma_z$) without or with SOC, along the $k$ path in the $\hat{\bf k}_y$ direction. Figures~\ref{fig_spin_textures}(a) and \ref{fig_spin_textures}(b) show the results at the Fermi energy, with Figs.~\ref{fig_spin_textures}(c) and \ref{fig_spin_textures}(d) providing an enlarged view of the regions near the $\Gamma$-point. In Figs.~\ref{fig_bands_projected}(d)--\ref{fig_bands_projected}(f) and \ref{fig_spin_textures}(e)--\ref{fig_spin_textures}(h) the colors indicate the proportion of contributions between the \ce{Cr3Te4} and \ce{PtTe2} layers.

As shown in Figs.~\ref{fig_bands_projected}(a) and \ref{fig_spin_textures}(a), in the absence of SOC, only Zeeman spin-splitting contributes to the $\sigma_y$ component. In Fig.~\ref{fig_spin_textures}(e), the hybridization between \ce{Cr3Te4} and \ce{PtTe2} (indicated in green) leads to the formation of two hybrid bands that are very close near the $\Gamma$-point. Outside of these hybrid bands, the spin-up (red) and spin-down (blue) states are separated due to the Zeeman spin-splitting caused by the magnetism of \ce{Cr3Te4}.
In contrast, when SOC is present, Rashba spin-splitting is observed on the hybrid bands. As shown in Fig.~\ref{fig_spin_textures}(c), $\sigma_x$ exhibits inversion symmetry along the $k_y$ direction, namely $\sigma_x (k_y)=-\sigma_x (-k_y)$, with nonzero $\sigma_x$ values across $k_y$, indicating a spin direction perpendicular to the electron momentum. A similar behavior is observed for $\sigma_y$ along the $k_x$ direction. These characteristics of spin-splitting confirm the presence of Rashba SOC induced by the proximity in the heterobilayer.
%

\section{Surface magnetic anisotropy}
\label{sec_sma}


In Fig.~\ref{fig_sma}, we present the surface magnetic anisotropy by computing the magnetic anisotropy energy (MAE, $\Delta E$) for different magnetization orientations: out-of-plane (from $[001]$ to $[100]$ or $[110]$) and in-plane (from $[110]$ to $[\bar{1}10]$). To ensure numerical accuracy, we increase the $k$-space mesh to $30\times30\times1$ and include a Hubbard $U=4.0$ eV on the Cr atoms in the VASP calculations. For accurate MAE calculations with VASP, we restored the full lattice symmetry for the GGA functional by setting \texttt{GGA\_COMPAT = .FALSE.}.
The results clearly show that the \ce{Cr3Te4}/\ce{PtTe2} bilayer has energy minima along the $[100]$ and $[110]$ directions, consistent with a $\sin^2\theta$ dependence, indicating uniaxial magnetic anisotropy.
The surface anisotropy energy is obtained from the GGA+$U$ calculations as $K_\mathrm{surf} = (\Delta E \cdot N_\mathrm{Cr})/A = -0.555$ $\mathrm{mJ}/\mathrm{m}^2$ with $\Delta E = E_{[110]} - E_{[001]} = -0.164$  $\mathrm{meV}/N_\mathrm{Cr}$, $N_\mathrm{Cr} = 3$, and unit cell area $A = 14.21$ {\AA}$^2$. The negative sign of $K_\mathrm{surf}$ indicates that \ce{Cr3Te4} has a hard axis along the $z$ direction.
For in-plane rotations from $[110]$ to $[\bar{1}10]$ direction, $\Delta E < \pm 0.01$ $\mathrm{meV}/N_\mathrm{Cr}$ is significantly smaller than in the out-of-plane case, indicating weak magnetocrystalline anisotropy within the $x$-$y$ plane.
Together, these results confirm the presence of an easy plane, characterizing the bilayer as an IMA system.

\section{Derivation of switching current}
\label{sec_derivation}

To derive the switching current for in-plane magnetic anisotropy (IMA) and perpendicular magnetic anisotropy (PMA) systems in the main text, we follow the procedures of \cite{Sun2000} and \cite{Xiao2005}, with an additional consideration for the contribution of fieldlike spin torque. For consistency, we align the easy axis along the $z$ axis and place the easy plane in the $z$-$y$ plane, such that the hard axis is along the $x$ axis. The unit vector of the magnetization is expressed as $\hat{\mathbf{m}} = (\sin\theta\cos\phi, \sin\theta\sin\phi, \cos\theta)$.
The uniaxial anisotropy energy is given by $U_\mathrm{uni} = -K_\mathrm{uni} m_z^2$, where $K_\mathrm{uni} = \frac{1}{2}\mu_0 M_\mathrm{s} H_\mathrm{k}$ and $H_\mathrm{k} > 0$ is the Stoner-Wohlfarth switching field. The easy-plane anisotropy energy is expressed as $U_\mathrm{plane} = -K_\mathrm{plane} m_x^2$, where $K_\mathrm{plane} = \frac{1}{2}\mu_0 M_\mathrm{s} H_\mathrm{p}$ and $H_\mathrm{p} = 2K_\mathrm{plane}/(\mu_0 M_\mathrm{s}) < 0$ is the field magnitude. The negative value of $K_\mathrm{plane} = K_\mathrm{demg} + K_\mathrm{surf}/t_\mathrm{FM}$ accounts for the demagnetization field of a magnetic thin film $K_\mathrm{demg} = -\frac{1}{2}\mu_0 M_\mathrm{s}^2$ and the surface anisotropy $K_\mathrm{surf}$. The Zeeman energy due to an external magnetic field is $U_\mathrm{ext} = -\mu_0\mathbf{M}\cdot\mathbf{H}_\mathrm{ext}$, where $\mathbf{H}_\mathrm{ext} = H_\mathrm{ext}(0, \sin\theta_h, \cos\theta_h)$ is applied in the $z$-$y$ plane by an angle $\theta_h$.

\begin{widetext}
We rewrite the Landau-Lifshitz-Gilbert (LLG) equation as
\begin{equation}
		\frac{1}{\gamma}\frac{d\hat{\mathbf{m}}}{dt}=-\hat{\mathbf{m}}\times \mathbf{B}_{\mathrm{eff}}+\frac{\alpha}{\gamma}\hat{\mathbf{m}}\times \frac{d\hat{\mathbf{m}}}{dt}
		-\hat{\mathbf{m}}\times \left( \hat{\mathbf{m}}\times B_{\mathrm{DL}}\hat{\mathbf{p}} \right) -\hat{\mathbf{m}}\times B_{\mathrm{FL}}\hat{\mathbf{p}} ,
\end{equation}
and the corresponding Landau-Lifshitz (LL) form is derived as
\begin{equation}\label{eq_LL}
		\frac{1}{\gamma ^{\prime}}\frac{d\hat{\mathbf{m}}}{dt}=-\hat{\mathbf{m}}\times \mathbf{B}_{\mathrm{eff}}-\hat{\mathbf{m}}\times \left( \hat{\mathbf{m}}\times \alpha \mathbf{B}_{\mathrm{eff}} \right)
		-\hat{\mathbf{m}}\times \left( B_{\mathrm{FL}}-\alpha B_{\mathrm{DL}} \right) \hat{\mathbf{p}}
		-\hat{\mathbf{m}}\times \left[ \hat{\mathbf{m}}\times \left( \alpha B_{\mathrm{FL}}+B_{\mathrm{DL}} \right) \hat{\mathbf{p}} \right] ,
\end{equation}
where we defined $\gamma ^{\prime}\equiv {\gamma}/{(1+\alpha ^2)}$.
The effective field is obtained by $\mathbf{B}_{\mathrm{eff}} = - \frac{1}{M} \frac{\partial U}{\partial\hat{\mathbf{m}}}$ with $U = U_\mathrm{uni} + U_\mathrm{plane} + U_\mathrm{ext}$, where the differential operator in spherical coordinate is $\frac{\partial}{\partial \hat{\mathbf{m}}}=\mathbf{e}_{\theta}\left( \frac{\partial}{\partial \theta} \right) +\mathbf{e}_{\phi}\left( \frac{1}{\sin \theta}\frac{\partial}{\partial \phi} \right)$. The current-induced spin torques are expressed in the form of field as $B_{\mathrm{FL}} = \mu_0\tau_\mathrm{FL}J_\mathrm{c}$ and $B_{\mathrm{DL}} = \mu_0\tau_\mathrm{DL}J_\mathrm{c}$ in the fieldlike and dampinglike directions, respectively. $\hat{\mathbf{p}}= (0, \sin\theta_p, \cos\theta_p)$ is the spin polarization direction of the spin current in the $z$-$y$ plane.
%
By deriving \eqref{eq_LL} in spherical coordinates, we first obtain a term from the uniaxial anisotropy as
\begin{equation}
	\left[ \begin{array}{c}
		d_t\theta _1\\
		d_t\phi _1\\
	\end{array} \right] =\gamma ^{\prime}\mu _0H_{\mathrm{k}}\left[ \begin{array}{c}
		-\alpha \sin \theta \cos \theta\\
		\cos \theta\\
	\end{array} \right] .
\end{equation}
The term from the easy-plane anisotropy is
\begin{equation}
	\left[ \begin{array}{c}
		d_t\theta _2\\
		d_t\phi _2\\
	\end{array} \right] =\gamma ^{\prime}\mu _0H_{\mathrm{p}}\left[ \begin{array}{c}
		\left( \alpha \cos \theta \cos \phi -\sin \phi \right) \sin \theta \cos \phi\\
		-\left( \cos \theta \cos \phi +\alpha \sin \phi \right) \cos \phi\\
	\end{array} \right] .
\end{equation}
The term from the external magnetic field is
\begin{equation}
	\left[ \begin{array}{c}
		d_t\theta _3\\
		d_t\phi _3\\
	\end{array} \right] =\gamma ^{\prime}\mu _0H_{\mathrm{ext}}\left[ \begin{array}{c}
		\alpha \cos \theta \sin \phi \sin \theta _h-\alpha \sin \theta \cos \theta _h+\cos \phi \sin \theta _h\\
		\left( -\cos \theta \sin \phi \sin \theta _h+\alpha \cos \phi \sin \theta _h+\sin \theta \cos \theta _h \right) /\sin \theta\\
	\end{array} \right] .
\end{equation}
The term from the dampinglike torque is
\begin{equation}
	\left[ \begin{array}{c}
		d_t\theta _4\\
		d_t\phi _4\\
	\end{array} \right] =\gamma ^{\prime}\mu _0\tau _{\mathrm{DL}}J_{\mathrm{c}}\left[ \begin{array}{c}
		\cos \theta \sin \phi \sin \theta _p-\sin \theta \cos \theta _p-\alpha \cos \phi \sin \theta _p\\
		\left( \alpha \cos \theta \sin \phi \sin \theta _p-\alpha \sin \theta \cos \theta _p+\cos \phi \sin \theta _p \right) /\sin \theta\\
	\end{array} \right] .
\end{equation}
The term from the fieldlike torque is
\begin{equation}
	\left[ \begin{array}{c}
		d_t\theta _5\\
		d_t\phi _5\\
	\end{array} \right] =\gamma ^{\prime}\mu _0\tau _{\mathrm{FL}}J_{\mathrm{c}}\left[ \begin{array}{c}
		\alpha \cos \theta \sin \phi \sin \theta _p-\alpha \sin \theta \cos \theta _p+\cos \phi \sin \theta _p\\
		\left( -\cos \theta \sin \phi \sin \theta _p+\sin \theta \cos \theta _p+\alpha \cos \phi \sin \theta _p \right) /\sin \theta\\
	\end{array} \right] .
\end{equation}
By applying small angle approximation ($\sin\theta\approx\theta$ and $\cos\approx1$), we have
\begin{equation}\label{eq_cmpqms}
	\begin{aligned}
		d_t\theta &=\gamma ^{\prime}\mu _0\theta \left[ -\alpha \left( H_{\mathrm{k}}+H_{\mathrm{ext}}+\tau _{\mathrm{FL}}J_{\mathrm{c}} \right) -\tau _{\mathrm{DL}}J_{\mathrm{c}}+H_{\mathrm{p}}\left( \alpha \cos \phi -\sin \phi \right) \cos \phi \right]
		\\
		d_t\phi &=\gamma ^{\prime}\mu _0\left[ H_{\mathrm{k}}+H_{\mathrm{ext}}+\tau _{\mathrm{FL}}J_{\mathrm{c}}-\alpha \tau _{\mathrm{DL}}J_{\mathrm{c}}-H_{\mathrm{p}}\left( \cos \phi +\alpha \sin \phi \right) \cos \phi \right] .
	\end{aligned}
\end{equation}
\end{widetext}

For the case of in-plane rotation in systems with IMA, if we assume that the easy-plane anisotropy is strong enough to confine the magnetization rotation in the $z$-$y$ plane, we arrive at the boundary condition $\phi=90^{\circ}$, which implies $d_t \phi = 0$. Additionally, by assuming $\theta_h = 0$ and $\theta_p=0$ (i.e. both the external magnetic field and spin polarization are aligned along the $z$ axis), we obtain the following relation:
\begin{equation}
	H_{\mathrm{k}}+H_{\mathrm{ext}}+\tau _{\mathrm{FL}}J_{\mathrm{c}}-\alpha \tau _{\mathrm{DL}}J_{\mathrm{c}}=0 .
\end{equation}
The switching condition occurs when the spin torques exceed the combined effect of the anisotropy and external fields, i.e. $\tau _{\mathrm{FL}}J_{\mathrm{sw}}-\alpha \tau _{\mathrm{DL}}J_{\mathrm{sw}} = H_{\mathrm{k}}+H_{\mathrm{ext}}$. Using the relation $\eta \equiv (2e/\hbar) M_\mathrm{s} t_\mathrm{FM} \mu_0 \tau$, the switching current can be written as
\begin{equation}\label{eq_conapc}
	J_{\mathrm{sw}}(\text{strong--IMA})=\frac{2e}{\hbar}\frac{\mu _0M_{\mathrm{s}}t_{\mathrm{FM}}}{\eta _{\mathrm{FL}}-\alpha \eta _{\mathrm{DL}}}\left( H_{\mathrm{k}}+H_{\mathrm{ext}} \right) .
\end{equation}

For the case of out-of-plane rotation in systems with PMA, the surface anisotropy overcomes the demagnetization field such that the perpendicular direction becomes the easy axis i.e., $K_\mathrm{uni} = \frac{1}{2}\mu_0 M_\mathrm{s} H_\mathrm{k} = K_\mathrm{demg} + K_\mathrm{surf}/t_\mathrm{FM} > 0$. The switching condition occurs when the velocity in the $\theta$ direction changes sign, which implies $d_t \theta = 0$. By removing the easy-plane anisotropy ($H_\mathrm{p} =0$) in \eqref{eq_cmpqms} and assuming $\theta_h = 0$ and $\theta_p=0$, we obtain the following relation:
\begin{equation}\label{eq_snoiqs}
	\alpha \left( H_{\mathrm{k}}+H_{\mathrm{ext}}+\tau _{\mathrm{FL}}J_{\mathrm{c}} \right) +\tau _{\mathrm{DL}}J_{\mathrm{c}} = 0 .
\end{equation}
The switching condition then becomes $\alpha \tau _{\mathrm{FL}}J_{\mathrm{sw}}+\tau _{\mathrm{DL}}J_{\mathrm{sw}} = \alpha H_{\mathrm{k}}+\alpha H_{\mathrm{ext}}$, leading to
\begin{equation}\label{eq_mjdwns}
	J_{\mathrm{sw}}(\mathrm{PMA})=\frac{2e}{\hbar}\frac{\mu _0M_{\mathrm{s}}t_{\mathrm{FM}}}{\eta _{\mathrm{FL}}+\frac{1}{\alpha}\eta _{\mathrm{DL}}}\left( H_{\mathrm{k}}+H_{\mathrm{ext}} \right) ,
\end{equation}
which matches the expression derived in Eq.~(29) of Ref.~\cite{Sun2000}, except for the additional $\eta _{\mathrm{FL}}$ in our expression.
Note that this equation can also be obtained by removing the easy-plane magnetic anisotropy ($H_\text{p}=0$) in Eq.~\eqref{eq_sjpsoq}, which is based on the method of energy stability analysis \cite{Sun2000, Bazaliy2004}.

\bibliography{bibliography}

@article{Alghamdi2019,
	title = {Highly Efficient Spin–Orbit Torque and Switching of Layered Ferromagnet {Fe$_3$GeTe$_2$}},
	volume = {19},
	ISSN = {1530-6992},
	_url = {http://dx.doi.org/10.1021/acs.nanolett.9b01043},
	DOI = {10.1021/acs.nanolett.9b01043},
	number = {7},
	journal = {Nano Letters},
	publisher = {American Chemical Society (ACS)},
	author = {Alghamdi,  Mohammed and Lohmann,  Mark and Li,  Junxue and Jothi,  Palani R. and Shao,  Qiming and Aldosary,  Mohammed and Su,  Tang and Fokwa,  Boniface P. T. and Shi,  Jing},
	year = {2019},
	month = jun,
	pages = {4400–4405}
}

@article{Amin2020,
	year = {2020},
	title = {{Interfacial spin–orbit torques}},
	author = {Amin, V. P. and Haney, P. M. and Stiles, M. D.},
	journal = {Journal of Applied Physics},
	issn = {0021-8979},
	doi = {10.1063/5.0024019},
	pmid = {34121763},
	pmcid = {PMC8194107},
	pages = {151101},
	number = {15},
	volume = {128}
}

@article{Bazaliy2004,
	title = {Current-induced magnetization switching in small domains of different anisotropies},
	author = {Bazaliy, Ya. B. and Jones, B. A. and Zhang, Shou-Cheng},
	_journal = {Phys. Rev. B},
	journal = {Physical Review B},
	volume = {69},
	issue = {9},
	pages = {094421},
	numpages = {19},
	year = {2004},
	month = {Mar},
	publisher = {American Physical Society},
	doi = {10.1103/PhysRevB.69.094421},
	_url = {https://link.aps.org/doi/10.1103/PhysRevB.69.094421}
}

@article{Cococcioni2005,
	title = {Linear response approach to the calculation of the effective interaction parameters in the $\mathrm{LDA}+\mathrm{U}$ method},
	author = {Cococcioni, Matteo and de Gironcoli, Stefano},
	_journal = {Phys. Rev. B},
	journal = {Physical Review B},
	volume = {71},
	issue = {3},
	pages = {035105},
	numpages = {16},
	year = {2005},
	month = {Jan},
	publisher = {American Physical Society},
	doi = {10.1103/PhysRevB.71.035105},
	url = {https://link.aps.org/doi/10.1103/PhysRevB.71.035105}
}

@article{Chua2021,
	author = {Chua, Rebekah and Zhou, Jun and Yu, Xiaojiang and Yu, Wei and Gou, Jian and Zhu, Rui and Zhang, Lei and Liu, Meizhuang and Breese, Mark B. H. and Chen, Wei and Loh, Kian Ping and Feng, Yuan Ping and Yang, Ming and Huang, Yu Li and Wee, Andrew T. S.},
	title = {Room Temperature Ferromagnetism of Monolayer Chromium Telluride with Perpendicular Magnetic Anisotropy},
	journal = {Advanced Materials},
	volume = {33},
	number = {42},
	pages = {2103360},
	keywords = {2D magnets, monolayer chromium telluride, room temperature ferromagnetism, scanning tunneling microscopy, X-ray magnetic circular dichroism},
	doi = {https://doi.org/10.1002/adma.202103360},
	_url = {https://advanced.onlinelibrary.wiley.com/doi/abs/10.1002/adma.202103360},
	year = {2021}
}

@book{Datta1995,
	doi={10.1017/CBO9780511805776},
	_url={https://doi.org/10.1017/CBO9780511805776},
	place={Cambridge},
	series={Cambridge Studies in Semiconductor Physics and Microelectronic Engineering},
	title={{Electronic Transport in Mesoscopic Systems}},
	publisher={Cambridge University Press, Cambridge, United Kingdom},
	author={Datta, Supriyo},
	year={1995},
	collection={Cambridge Studies in Semiconductor Physics and Microelectronic Engineering}
}

@article{Deng2018,
	title = {Gate-tunable room-temperature ferromagnetism in two-dimensional {Fe$_3$GeTe$_2$}},
	volume = {563},
	ISSN = {1476-4687},
	_url = {http://dx.doi.org/10.1038/s41586-018-0626-9},
	DOI = {10.1038/s41586-018-0626-9},
	number = {7729},
	journal = {Nature},
	publisher = {Springer Science and Business Media LLC},
	author = {Deng,  Yujun and Yu,  Yijun and Song,  Yichen and Zhang,  Jingzhao and Wang,  Nai Zhou and Sun,  Zeyuan and Yi,  Yangfan and Wu,  Yi Zheng and Wu,  Shiwei and Zhu,  Junyi and Wang,  Jing and Chen,  Xian Hui and Zhang,  Yuanbo},
	year = {2018},
	month = oct,
	pages = {94–99}
}

@article{Dolui2020,
	title = {Proximity Spin–Orbit Torque on a Two-Dimensional Magnet within van der {Waals} Heterostructure: Current-Driven Antiferromagnet-to-Ferromagnet Reversible Nonequilibrium Phase Transition in Bilayer {CrI$_3$}},
	volume = {20},
	ISSN = {1530-6992},
	_url = {http://dx.doi.org/10.1021/acs.nanolett.9b04556},
	DOI = {10.1021/acs.nanolett.9b04556},
	number = {4},
	journal = {Nano Letters},
	publisher = {American Chemical Society (ACS)},
	author = {Dolui,  Kapildeb and Petrović,  Marko D. and Zollner,  Klaus and Plecháč,  Petr and Fabian,  Jaroslav and Nikolić,  Branislav K.},
	year = {2020},
	month = mar,
	pages = {2288–2295}
}

@article{Edelstein1990,
	title = {Spin polarization of conduction electrons induced by electric current in two-dimensional asymmetric electron systems},
	volume = {73},
	ISSN = {0038-1098},
	url = {http://dx.doi.org/10.1016/0038-1098(90)90963-C},
	DOI = {10.1016/0038-1098(90)90963-c},
	number = {3},
	journal = {Solid State Communications},
	publisher = {Elsevier BV},
	author = {Edelstein,  V.M.},
	year = {1990},
	month = jan,
	pages = {233–235}
}

@article{Fei2018,
	title = {Two-dimensional itinerant ferromagnetism in atomically thin {Fe$_3$GeTe$_2$}},
	volume = {17},
	ISSN = {1476-4660},
	_url = {http://dx.doi.org/10.1038/s41563-018-0149-7},
	DOI = {10.1038/s41563-018-0149-7},
	number = {9},
	journal = {Nature Materials},
	publisher = {Springer Science and Business Media LLC},
	author = {Fei,  Zaiyao and Huang,  Bevin and Malinowski,  Paul and Wang,  Wenbo and Song,  Tiancheng and Sanchez,  Joshua and Yao,  Wang and Xiao,  Di and Zhu,  Xiaoyang and May,  Andrew F. and Wu,  Weida and Cobden,  David H. and Chu,  Jiun-Haw and Xu,  Xiaodong},
	year = {2018},
	month = aug,
	pages = {778–782}
}

@article{Fukami2016,
	title = {A spin–orbit torque switching scheme with collinear magnetic easy axis and current configuration},
	volume = {11},
	ISSN = {1748-3395},
	url = {http://dx.doi.org/10.1038/NNANO.2016.29},
	DOI = {10.1038/nnano.2016.29},
	number = {7},
	journal = {Nature Nanotechnology},
	publisher = {Springer Science and Business Media LLC},
	author = {Fukami,  S. and Anekawa,  T. and Zhang,  C. and Ohno,  H.},
	year = {2016},
	month = mar,
	pages = {621–625}
}

@article{Freimuth2014,
	title = {Spin-orbit torques in Co/Pt(111) and Mn/W(001) magnetic bilayers from first principles},
	author = {Freimuth, Frank and Bl\"ugel, Stefan and Mokrousov, Yuriy},
	journal = {Physical Review B},
	volume = {90},
	issue = {17},
	pages = {174423},
	numpages = {10},
	year = {2014},
	month = {Nov},
	publisher = {American Physical Society},
	doi = {10.1103/PhysRevB.90.174423},
	url = {https://link.aps.org/doi/10.1103/PhysRevB.90.174423}
}

@article{Garello2013,
	title = {Symmetry and magnitude of spin–orbit torques in ferromagnetic heterostructures},
	volume = {8},
	ISSN = {1748-3395},
	_url = {http://dx.doi.org/10.1038/nnano.2013.145},
	DOI = {10.1038/nnano.2013.145},
	number = {8},
	journal = {Nature Nanotechnology},
	publisher = {Springer Science and Business Media LLC},
	author = {Garello,  Kevin and Miron,  Ioan Mihai and Avci,  Can Onur and Freimuth,  Frank and Mokrousov,  Yuriy and Bl\"{u}gel,  Stefan and Auffret,  Stéphane and Boulle,  Olivier and Gaudin,  Gilles and Gambardella,  Pietro},
	year = {2013},
	month = jul,
	pages = {587–593}
}

@article{Go2020,
	doi = {10.1103/physrevresearch.2.033401},
	_url = {https://doi.org/10.1103/physrevresearch.2.033401},
	year = {2020},
	month = sep,
	publisher = {American Physical Society ({APS})},
	volume = {2},
	number = {3},
	author = {Dongwook Go and Frank Freimuth and Jan-Philipp Hanke and Fei Xue and Olena Gomonay and Kyung-Jin Lee and Stefan Bl\"{u}gel and Paul M. Haney and Hyun-Woo Lee and Yuriy Mokrousov},
	title = {Theory of current-induced angular momentum transfer dynamics in spin-orbit coupled systems},
	journal = {Physical Review Research},
	pages = {033401}
}

@article{Goswami2024,
	title = {Critical behavior in monoclinic {Cr$_3$Te$_4$}},
	author = {Goswami, Anirban and Ng, Nicholas and Yakubu, Emmanuel and Abeykoon, AM Milinda and Guchhait, Samaresh},
	_journal = {Phys. Rev. B},
	journal = {Physical Review B},
	volume = {109},
	issue = {5},
	pages = {054413},
	numpages = {10},
	year = {2024},
	month = {Feb},
	publisher = {American Physical Society},
	doi = {10.1103/PhysRevB.109.054413},
	url = {https://link.aps.org/doi/10.1103/PhysRevB.109.054413}
}

@article{Haney2013,
	title = {Current-induced torques and interfacial spin-orbit coupling},
	volume = {88},
	ISSN = {1550-235X},
	_url = {http://dx.doi.org/10.1103/physrevb.88.214417},
	DOI = {10.1103/physrevb.88.214417},
	number = {21},
	journal = {Physical Review B},
	publisher = {American Physical Society (APS)},
	author = {Haney,  Paul M. and Lee,  Hyun-Woo and Lee,  Kyung-Jin and Manchon,  Aurélien and Stiles,  M. D.},
	year = {2013},
	month = dec,
	pages = {214417}
}

@article{Hideki2024,
	author = {Hideki Matsuoka  and Shun Kajihara  and Takuya Nomoto  and Yue Wang  and Motoaki Hirayama  and Ryotaro Arita  and Yoshihiro Iwasa  and Masaki Nakano },
	title = {Band-driven switching of magnetism in a van der Waals magnetic semimetal},
	journal = {Science Advances},
	volume = {10},
	number = {15},
	pages = {eadk1415},
	year = {2024},
	doi = {10.1126/sciadv.adk1415},
	URL = {https://www.science.org/doi/abs/10.1126/sciadv.adk1415}
}

@article{Huang2017,
	title = {Layer-dependent ferromagnetism in a van der Waals crystal down to the monolayer limit},
	volume = {546},
	ISSN = {1476-4687},
	_url = {http://dx.doi.org/10.1038/nature22391},
	DOI = {10.1038/nature22391},
	number = {7657},
	journal = {Nature},
	publisher = {Springer Science and Business Media LLC},
	author = {Huang,  Bevin and Clark,  Genevieve and Navarro-Moratalla,  Efrén and Klein,  Dahlia R. and Cheng,  Ran and Seyler,  Kyle L. and Zhong,  Ding and Schmidgall,  Emma and McGuire,  Michael A. and Cobden,  David H. and Yao,  Wang and Xiao,  Di and Jarillo-Herrero,  Pablo and Xu,  Xiaodong},
	year = {2017},
	month = jun,
	pages = {270–273}
}

@article{Huang2023a,
	title = {{Validity of DFT-based spin-orbit torque calculation for perpendicular magnetic anisotropy in iron thin films}},
	volume = {13},
	ISSN = {2158-3226},
	_url = {http://dx.doi.org/10.1063/9.0000481},
	DOI = {10.1063/9.0000481},
	number = {1},
	journal = {AIP Advances},
	publisher = {AIP Publishing},
	author = {Huang,  Bao-Huei and Lai,  Yi-Feng and Tang,  Yu-Hui},
	year = {2023},
	month = jan,
	pages = {015034}
}

@article{Huang2023b,
	title = {{Determining perpendicular magnetic anisotropy in Fe/MgO/Fe magnetic tunnel junction: A DFT-based spin–orbit torque method}},
	volume = {585},
	ISSN = {0304-8853},
	_url = {http://dx.doi.org/10.1016/j.jmmm.2023.171098},
	DOI = {10.1016/j.jmmm.2023.171098},
	journal = {Journal of Magnetism and Magnetic Materials},
	publisher = {Elsevier BV},
	author = {Huang,  Bao-Huei and Fu,  Yu-Hsiang and Kaun,  Chao-Cheng and Tang,  Yu-Hui},
	year = {2023},
	month = nov,
	pages = {171098}
}

@article{Ke2008,
	title = {Disorder Scattering in Magnetic Tunnel Junctions: Theory of Nonequilibrium Vertex Correction},
	author = {Ke, Youqi and Xia, Ke and Guo, Hong},
	_journal = {Phys. Rev. Lett.},
	journal = {Physical Review Letters},
	volume = {100},
	issue = {16},
	pages = {166805},
	numpages = {4},
	year = {2008},
	month = {Apr},
	publisher = {American Physical Society},
	doi = {10.1103/PhysRevLett.100.166805},
	_url = {https://link.aps.org/doi/10.1103/PhysRevLett.100.166805},
}

@article{Klime2011,
  title = {Van der Waals density functionals applied to solids},
  author = {Klime\ifmmode \check{s}\else \v{s}\fi{}, Ji\ifmmode \check{r}\else \v{r}\fi{}\'{\i} and Bowler, David R. and Michaelides, Angelos},
  journal = {Physical Review B},
  volume = {83},
  issue = {19},
  pages = {195131},
  numpages = {13},
  year = {2011},
  month = {May},
  publisher = {American Physical Society},
  doi = {10.1103/PhysRevB.83.195131},
  url = {https://link.aps.org/doi/10.1103/PhysRevB.83.195131}
}

@article{Kresse1996,
	doi = {10.1103/physrevb.54.11169},
	_url = {https://doi.org/10.1103/physrevb.54.11169},
	year = {1996},
	month = oct,
	publisher = {American Physical Society ({APS})},
	volume = {54},
	number = {16},
	pages = {11169--11186},
	author = {G. Kresse and J. Furthm\"{u}ller},
	title = {Efficient iterative schemes for \textit{ab initio} total-energy calculations using a plane-wave basis set},
	journal = {Physical Review B},
	_journal = {Phys. Rev. B}
}

@article{Kresse1999,
	doi = {10.1103/physrevb.59.1758},
	_url = {https://doi.org/10.1103/physrevb.59.1758},
	year = {1999},
	month = jan,
	publisher = {American Physical Society ({APS})},
	volume = {59},
	number = {3},
	pages = {1758--1775},
	author = {G. Kresse and D. Joubert},
	title = {From ultrasoft pseudopotentials to the projector augmented-wave method},
	journal = {Physical Review B},
	_journal = {Phys. Rev. B}
}

@article{Kim2017,
	title = {Spin-orbit torques from interfacial spin-orbit coupling for various interfaces},
	volume = {96},
	ISSN = {2469-9969},
	_url = {http://dx.doi.org/10.1103/PhysRevB.96.104438},
	DOI = {10.1103/physrevb.96.104438},
	number = {10},
	journal = {Physical Review B},
	publisher = {American Physical Society (APS)},
	author = {Kim,  Kyoung-Whan and Lee,  Kyung-Jin and Sinova,  Jairo and Lee,  Hyun-Woo and Stiles,  M. D.},
	year = {2017},
	month = sep,
	pages = {104438}
}

@article{Lee2015,
	year = {2015},
	title = {{Angular dependence of spin-orbit spin-transfer torques}},
	author = {Lee, Ki-Seung and Go, Dongwook and Manchon, Aurélien and Haney, Paul M. and Stiles, M. D. and Lee, Hyun-Woo and Lee, Kyung-Jin},
	journal = {Physical Review B},
	issn = {1098-0121},
	doi = {10.1103/physrevb.91.144401},
	pages = {144401},
	number = {14},
	volume = {91}
}

@article{Li2019,
	title = {Molecular Beam Epitaxy Grown {Cr$_2$Te$_3$} Thin Films with Tunable Curie Temperatures for Spintronic Devices},
	volume = {2},
	ISSN = {2574-0970},
	_url = {http://dx.doi.org/10.1021/acsanm.9b01179},
	DOI = {10.1021/acsanm.9b01179},
	number = {11},
	journal = {ACS Applied Nano Materials},
	publisher = {American Chemical Society (ACS)},
	author = {Li,  Hongxi and Wang,  Linjing and Chen,  Junshu and Yu,  Tao and Zhou,  Liang and Qiu,  Yang and He,  Hongtao and Ye,  Fei and Sou,  Iam Keong and Wang,  Gan},
	year = {2019},
	month = oct,
	pages = {6809–6817}
}

@article{Liu2012,
	title = {Spin-Torque Switching with the Giant Spin Hall Effect of Tantalum},
	volume = {336},
	ISSN = {1095-9203},
	_url = {http://dx.doi.org/10.1126/science.1218197},
	DOI = {10.1126/science.1218197},
	number = {6081},
	journal = {Science},
	publisher = {American Association for the Advancement of Science (AAAS)},
	author = {Liu,  Luqiao and Pai,  Chi-Feng and Li,  Y. and Tseng,  H. W. and Ralph,  D. C. and Buhrman,  R. A.},
	year = {2012},
	month = may,
	pages = {555–558}
}

@article{Liu2012Ralph,
	year = {2012},
	title = {{Current-Induced Switching of Perpendicularly Magnetized Magnetic Layers Using Spin Torque from the Spin Hall Effect}},
	author = {Liu, Luqiao and Lee, O. J. and Gudmundsen, T. J. and Ralph, D. C. and Buhrman, R. A.},
	journal = {Physical Review Letters},
	issn = {0031-9007},
	doi = {10.1103/physrevlett.109.096602},
	pmid = {23002867},
	number = {9},
	volume = {109},
	pages = {096602}
}

@article{Lv2018,
	title = {Electric-Field Control of Spin–Orbit Torques in {WS$_2$}/Permalloy Bilayers},
	volume = {10},
	ISSN = {1944-8252},
	_url = {http://dx.doi.org/10.1021/acsami.7b16919},
	DOI = {10.1021/acsami.7b16919},
	number = {3},
	journal = {ACS Applied Materials Interfaces},
	publisher = {American Chemical Society (ACS)},
	author = {Lv,  Weiming and Jia,  Zhiyan and Wang,  Bochong and Lu,  Yuan and Luo,  Xin and Zhang,  Baoshun and Zeng,  Zhongming and Liu,  Zhongyuan},
	year = {2018},
	month = jan,
	pages = {2843–2849}
}

@article{Mahfouzi2020,
	title = {Microscopic origin of spin-orbit torque in ferromagnetic heterostructures: A first-principles approach},
	volume = {101},
	ISSN = {2469-9969},
	_url = {http://dx.doi.org/10.1103/PhysRevB.101.060405},
	DOI = {10.1103/physrevb.101.060405},
	number = {6},
	journal = {Physical Review B},
	publisher = {American Physical Society (APS)},
	author = {Mahfouzi,  Farzad and Mishra,  Rahul and Chang,  Po-Hao and Yang,  Hyunsoo and Kioussis,  Nicholas},
	year = {2020},
	month = feb,
	pages = {060405}
}

@article{Manchon2019,
	title = {Current-induced spin-orbit torques in ferromagnetic and antiferromagnetic systems},
	author = {Manchon, A. and \ifmmode \check{Z}\else \v{Z}\fi{}elezn\'y, J. and Miron, I. M. and Jungwirth, T. and Sinova, J. and Thiaville, A. and Garello, K. and Gambardella, P.},
	journal = {Reviews of Modern Physics},
	volume = {91},
	issue = {3},
	pages = {035004},
	numpages = {80},
	year = {2019},
	month = {Sep},
	publisher = {American Physical Society},
	doi = {10.1103/RevModPhys.91.035004},
	url = {https://link.aps.org/doi/10.1103/RevModPhys.91.035004}
}

@article{Miron2011,
	year = {2011},
	title = {{Perpendicular switching of a single ferromagnetic layer induced by in-plane current injection}},
	author = {Miron, Ioan Mihai and Garello, Kevin and Gaudin, Gilles and Zermatten, Pierre-Jean and Costache, Marius V. and Auffret, Stéphane and Bandiera, Sébastien and Rodmacq, Bernard and Schuhl, Alain and Gambardella, Pietro},
	journal = {Nature},
	issn = {0028-0836},
	doi = {10.1038/nature10309},
	pmid = {21804568},
	pages = {189--193},
	number = {7359},
	volume = {476}
}

@article{Nakayama2016,
	year = {2016},
	title = {{Rashba-Edelstein Magnetoresistance in Metallic Heterostructures}},
	author = {Nakayama, Hiroyasu and Kanno, Yusuke and An, Hongyu and Tashiro, Takaharu and Haku, Satoshi and Nomura, Akiyo and Ando, Kazuya},
	journal = {Physical Review Letters},
	issn = {0031-9007},
	doi = {10.1103/physrevlett.117.116602},
	pmid = {27661708},
	pages = {116602},
	number = {11},
	volume = {117}
}

@article{Perdew1981,
  title = {Self-interaction correction to density-functional approximations for many-electron systems},
  author = {Perdew, J. P. and Zunger, Alex},
  journal = {Physical Review B},
  volume = {23},
  issue = {10},
  pages = {5048--5079},
  numpages = {0},
  year = {1981},
  month = {May},
  publisher = {American Physical Society},
  doi = {10.1103/PhysRevB.23.5048},
  url = {https://link.aps.org/doi/10.1103/PhysRevB.23.5048}
}

@article{Perdew1996,
	title = {Generalized Gradient Approximation Made Simple},
	volume = {77},
	ISSN = {1079-7114},
	_url = {http://dx.doi.org/10.1103/PhysRevLett.77.3865},
	DOI = {10.1103/physrevlett.77.3865},
	number = {18},
	journal = {Physical Review Letters},
	publisher = {American Physical Society (APS)},
	author = {Perdew,  John P. and Burke,  Kieron and Ernzerhof,  Matthias},
	year = {1996},
	month = oct,
	pages = {3865–3868}
}

@article{Ramaswamy2018,
	title = {Recent advances in spin-orbit torques: Moving towards device applications},
	volume = {5},
	ISSN = {1931-9401},
	url = {http://dx.doi.org/10.1063/1.5041793},
	DOI = {10.1063/1.5041793},
	number = {3},
	journal = {Applied Physics Reviews},
	publisher = {AIP Publishing},
	author = {Ramaswamy,  Rajagopalan and Lee,  Jong Min and Cai,  Kaiming and Yang,  Hyunsoo},
	year = {2018},
	month = sep,
	pages = {031107}
}

@article{Shin2022,
	title = {Spin–Orbit Torque Switching in an All‐Van der {Waals} Heterostructure},
	volume = {34},
	ISSN = {1521-4095},
	_url = {http://dx.doi.org/10.1002/adma.202101730},
	DOI = {10.1002/adma.202101730},
	number = {8},
	journal = {Advanced Materials},
	publisher = {Wiley},
	author = {Shin,  Inseob and Cho,  Won Joon and An,  Eun‐Su and Park,  Sungyu and Jeong,  Hyeon‐Woo and Jang,  Seong and Baek,  Woon Joong and Park,  Seong Yong and Yang,  Dong‐Hwan and Seo,  Jun Ho and Kim,  Gi‐Yeop and Ali,  Mazhar N. and Choi,  Si‐Young and Lee,  Hyun‐Woo and Kim,  Jun Sung and Kim,  Sung Dug and Lee,  Gil‐Ho},
	year = {2022},
	month = jan,
	pages = {2101730}
}

@article{Sinova2004,
	year = {2004},
	title = {{Universal Intrinsic Spin Hall Effect}},
	author = {Sinova, Jairo and Culcer, Dimitrie and Niu, Q. and Sinitsyn, N. A. and Jungwirth, T. and MacDonald, A. H.},
	journal = {Physical Review Letters},
	issn = {0031-9007},
	doi = {10.1103/physrevlett.92.126603},
	pmid = {15089695},
	pages = {126603},
	number = {12},
	volume = {92},
}

@article{Sinova2015,
	title = {Spin Hall effects},
	volume = {87},
	ISSN = {1539-0756},
	url = {http://dx.doi.org/10.1103/RevModPhys.87.1213},
	DOI = {10.1103/revmodphys.87.1213},
	number = {4},
	journal = {Reviews of Modern Physics},
	publisher = {American Physical Society (APS)},
	author = {Sinova,  Jairo and Valenzuela,  Sergio O. and Wunderlich,  J. and Back,  C. H. and Jungwirth,  T.},
	year = {2015},
	month = oct,
	pages = {1213–1260}
}

@article{Sun2000,
	year = {2000},
	rating = {5},
	title = {{Spin-current interaction with a monodomain magnetic body: A model study}},
	author = {Sun, J. Z.},
	_journal = {Phys. Rev. B},
	journal = {Physical Review B},
	issn = {1098-0121},
	doi = {10.1103/physrevb.62.570},
	pages = {570--578},
	number = {1},
	volume = {62},
	keywords = {},
}

@article{Taylor2001,
	doi = {10.1103/physrevb.63.245407},
	_url = {https://doi.org/10.1103/physrevb.63.245407},
	year = {2001},
	month = jun,
	publisher = {American Physical Society ({APS})},
	volume = {63},
	number = {24},
	author = {Jeremy Taylor and Hong Guo and Jian Wang},
	title = {\textit{Ab initio} modeling of quantum transport properties of molecular electronic devices},
	journal = {Physical Review B},
	pages = {245407}
}

@article{Tian2021,
	title = {Two-Dimensional Van Der Waals Materials for Spin-Orbit Torque Applications},
	volume = {3},
	number = {732916},
	pages = {1--12},
	ISSN = {2673-3013},
	_url = {http://dx.doi.org/10.3389/fnano.2021.732916},
	DOI = {10.3389/fnano.2021.732916},
	journal = {Frontiers in Nanotechnology},
	publisher = {Frontiers Media SA},
	author = {Tian,  Mingming and Zhu,  Yonghui and Jalali,  Milad and Jiang,  Wei and Liang,  Jian and Huang,  Zhaocong and Chen,  Qian and Zeng,  Zhongming and Zhai,  Ya},
	year = {2021},
	month = aug
}

@article{Timopheev2015,
	year = {2015},
	title = {{Respective influence of in-plane and out-of-plane spin-transfer torques in magnetization switching of perpendicular magnetic tunnel junctions}},
	author = {Timopheev, A. A. and Sousa, R. and Chshiev, M. and Buda-Prejbeanu, L. D. and Dieny, B.},
	journal = {Physical Review B},
	issn = {1098-0121},
	doi = {10.1103/physrevb.92.104430},
	pages = {104430},
	number = {10},
	volume = {92}
}

@article{Waldron2007,
	doi = {10.1088/0957-4484/18/42/424026},
	_url = {https://doi.org/10.1088/0957-4484/18/42/424026},
	year = {2007},
	month = sep,
	publisher = {{IOP} Publishing},
	volume = {18},
	number = {42},
	pages = {424026},
	author = {Derek Waldron and Lei Liu and Hong Guo},
	title = {\textit{Ab initio} simulation of magnetic tunnel junctions},
	journal = {Nanotechnology}
}

@article{Xiao2005,
	year = {2005},
	title = {{Macrospin models of spin transfer dynamics}},
	author = {Xiao, Jiang and Zangwill, A. and Stiles, M. D.},
	journal = {Physical Review B},
	issn = {1098-0121},
	doi = {10.1103/physrevb.72.014446},
	journaltitle = {Physical Review B},
	pages = {014446},
	number = {1},
	volume = {72}
}

@article{Xu2020,
	title = {High Spin {Hall} Conductivity in Large‐Area Type‐{II} {Dirac} Semimetal {PtTe$_2$}},
	volume = {32},
	ISSN = {1521-4095},
	_url = {http://dx.doi.org/10.1002/adma.202000513},
	DOI = {10.1002/adma.202000513},
	number = {17},
	journal = {Advanced Materials},
	publisher = {Wiley},
	author = {Xu,  Hongjun and Wei,  Jinwu and Zhou,  Hengan and Feng,  Jiafeng and Xu,  Teng and Du,  Haifeng and He,  Congli and Huang,  Yuan and Zhang,  Junwei and Liu,  Yizhou and Wu,  Han‐Chun and Guo,  Chenyang and Wang,  Xiao and Guang,  Yao and Wei,  Hongxiang and Peng,  Yong and Jiang,  Wanjun and Yu,  Guoqiang and Han,  Xiufeng},
	year = {2020},
	month = mar,
	pages = {2000513}
}

@article{Yamaguchi1972,
	title = {Magnetic Properties of {Cr$_3$Te$_4$} in Ferromagnetic Region},
	volume = {32},
	ISSN = {1347-4073},
	_url = {http://dx.doi.org/10.1143/jpsj.32.635},
	DOI = {10.1143/jpsj.32.635},
	number = {3},
	journal = {Journal of the Physical Society of Japan},
	publisher = {Physical Society of Japan},
	author = {Yamaguchi,  Masuhiro and Hashimoto,  Takasu},
	year = {1972},
	month = mar,
	pages = {635–638}
}

@article{Zhu2018,
	title = {Systematic search for two-dimensional ferromagnetic materials},
	volume = {2},
	ISSN = {2475-9953},
	_url = {http://dx.doi.org/10.1103/PhysRevMaterials.2.081001},
	DOI = {10.1103/physrevmaterials.2.081001},
	number = {8},
	journal = {Physical Review Materials},
	publisher = {American Physical Society (APS)},
	author = {Zhu,  Yu and Kong,  Xianghua and Rhone,  Trevor David and Guo,  Hong},
	year = {2018},
	month = aug,
	pages = {081001}
}

\end{document}